# Information Theory and Statistical Learning

**Abbas El Gamal**

Stanford University

# CHAPTER 1

# Statistical Learning

Statistical learning—often considered a subfield of, or synonymous with, *machine learning*—is concerned with training probabilistic models from data. These models are then used for *prediction*, *sample generation*, and the discovery of compact, meaningful *representations* of the data.

Information theory intersects with statistical learning in two principal ways: through the objective functions used to train probabilistic models, and through the fundamental limits that govern the performance of learning algorithms. This chapter explores the former connection; the next addresses the latter. Although optimization algorithms and implementation details are essential for the practical success of statistical learning, our focus here is on how information-theoretic concepts arise in the training of a broad range of statistical learning models.

The central problem in statistical learning is simple to state (Figure 1.1): given a dataset sampled from an unknown distribution $p_{\text{data}}$ and a parametric family of distributions (also called a *probability model*) $\mathscr{P} = \{p_\theta : \theta \in \Theta\}$, identify the distribution $p_{\hat{\theta}} \in \mathscr{P}$ that is closest to $p_{\text{data}}$. This problem is conceptually related to parameter estimation, where one generally assumes the existence of a parameter $\theta^* \in \Theta$ such that $p_{\theta^*} = p_{\text{data}}$, and the goal is to estimate $\theta^*$ from the data. The statistical learning problem departs from this classical paradigm in a fundamental way: we do not assume that $p_{\text{data}} \in \mathscr{P}$. Indeed, the true data-generating distribution may lie outside $\mathscr{P}$ entirely. In this view, the choice of $\mathscr{P}$ reflects prior knowledge, modeling assumptions, or inductive biases about the data-generating process.

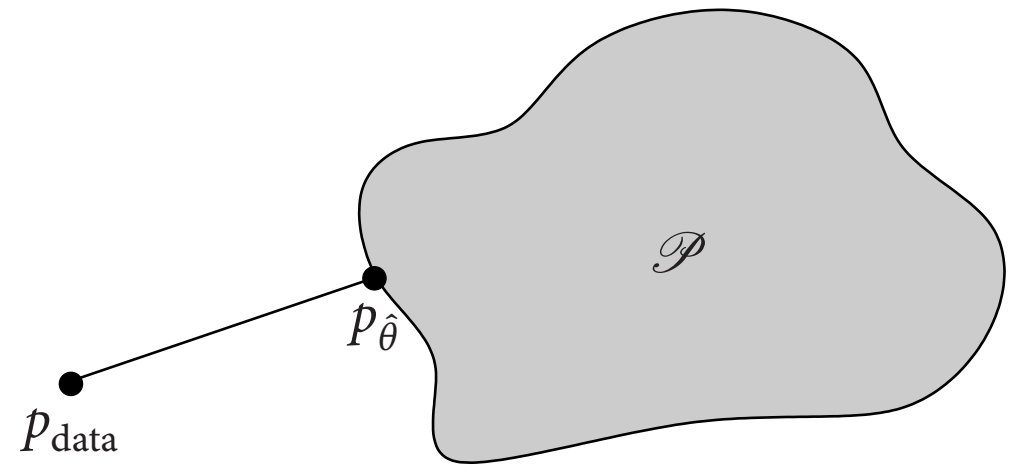

**Figure 1.1.** Statistical learning problem.

The first step in formulating a learning problem is to specify a probability model together with an appropriate measure of divergence between probability distributions. Among such measures, relative entropy is the most widely used, in part because minimizing it is equivalent to maximizing the expected log-likelihood under the unknown data-generating distribution. In some settings, however, alternative information-theoretic divergences are employed. Because the data-generating distribution is unknown, these ideal learning objectives cannot be evaluated directly; instead, they are approximated by their empirical counterparts computed from the observed data. Solving the resulting empirical optimization problem—a process known as *model training*—yields a *learned model* $p_{\hat{\theta}}$. For some models, however, direct maximization of the empirical log-likelihood is computationally intractable, and one instead maximizes a suitable lower bound.

We begin with *supervised learning*, in which a model is learned from labeled data, and illustrate it through the classical predictive models of linear and logistic regression. This discussion serves as a prelude to neural networks, which underlie many of the models discussed later in the chapter. We then turn to *generative models*, whose primary goal is to produce samples resembling those in the training dataset. We discuss autoregressive models, such as large language models (LLMs); variational autoencoders (VAEs) and diffusion models, trained by maximizing the evidence lower bound (ELBO) on the log-likelihood (Section 1.4.2); and score-based models, trained using the Fisher divergence (Section 1.7.1).

A central tension underlies statistical learning: a good model must balance fitting the training data against generalizing to unseen samples. An overfitted model may achieve excellent performance on the training set yet perform poorly on new data. This trade-off is often addressed through *regularization*, which limits the effective complexity of the learned model by penalizing overly complex solutions. In practice, generalization performance is typically assessed by partitioning the available data into a *training set* and a *test set*, and measuring the learned model's performance on the latter. We develop bounds on generalization error in the next chapter.

**Notation.** In this chapter, we denote datasets by $\{\mathbf{x}_i\}_{i=1}^n$ or, for input–output data, by $\{(\mathbf{x}_i, \mathbf{y}_i)\}_{i=1}^n$, where, unless stated otherwise, the pairs $(\mathbf{x}_i, \mathbf{y}_i)$ take values in $\mathbb{R}^K \times \mathbb{R}^M$. We use natural logarithms rather than base-2 logarithms throughout and occasionally use the subscript NN to denote a function or probability distribution modeled by a neural network.

## 1.1 SUPERVISED LEARNING

For concreteness, in the following, we consider discrete random variables. The same discussion applies to continuous random variables with $p$ replaced with $f$, and entropy replaced with differential entropy.

In supervised learning, we are given a conditional probability model $\mathscr{P} = \{p_\theta(\mathbf{y}|\mathbf{x}) : \theta \in \Theta\}$ that describes how the output (or *label*) $\mathbf{Y}$ is randomly generated given an input (or *feature vector*) $\mathbf{X}$, and a dataset of input–output pairs $\{(\mathbf{x}_i, \mathbf{y}_i)\}_{i=1}^n$ drawn i.i.d. according to

an unknown joint pmf $p_{\text{data}}(\mathbf{x}, \mathbf{y})$. The objective is to find a conditional pmf $p_{\hat{\theta}}(\mathbf{y}|\mathbf{x}) \in \mathscr{P}$ that is closest to the true conditional pmf $p_{\text{data}}(\mathbf{y}|\mathbf{x})$ in conditional relative entropy.

We first show that this objective is equivalent to minimizing the conditional cross entropy, or equivalently, maximizing the expected conditional log-likelihood.

**Lemma 1.1.** Assume that $H(\mathbf{Y}|\mathbf{X}) < \infty$. Then minimizing the conditional relative entropy $D\big(p_{\text{data}}(\mathbf{y}|\mathbf{x}) \| p_\theta(\mathbf{y}|\mathbf{x}) | p_{\text{data}}(\mathbf{x})\big)$ over $\theta$ is equivalent to minimizing the conditional cross entropy $H\big(p_{\text{data}}(\mathbf{y}|\mathbf{x}), p_\theta(\mathbf{y}|\mathbf{x}) | p_{\text{data}}(\mathbf{x})\big)$.

**Proof.** We expand the conditional relative entropy:

$$D\big(p_{\text{data}}(\mathbf{y}|\mathbf{x}) \| p_\theta(\mathbf{y}|\mathbf{x}) | p_{\text{data}}(\mathbf{x})\big) = \mathsf{E}_{p_{\text{data}}(\mathbf{x},\mathbf{y})}\big(\ln p_{\text{data}}(\mathbf{Y}|\mathbf{X})\big) - \mathsf{E}_{p_{\text{data}}(\mathbf{x},\mathbf{y})}\big(\ln p_\theta(\mathbf{Y}|\mathbf{X})\big).$$

The first term is the negative conditional entropy $-H(\mathbf{Y}|\mathbf{X})$ under the data-generating distribution and is not a function of $\theta$. Hence, minimizing the conditional relative entropy is the same as minimizing

$$-\mathsf{E}_{p_{\text{data}}(\mathbf{x},\mathbf{y})}\big(\ln p_\theta(\mathbf{Y}|\mathbf{X})\big), \tag{1.1}$$

which is exactly the conditional cross entropy $H\big(p_{\text{data}}(\mathbf{y}|\mathbf{x}), p_\theta(\mathbf{y}|\mathbf{x}) | p_{\text{data}}(\mathbf{x})\big)$.

Since the true distribution of the data is unknown, we instead maximize the conditional log-likelihood of the data.

**Definition 1.1.** Given a dataset $\{(\mathbf{x}_i, \mathbf{y}_i)\}_{i=1}^n$, we define the conditional log-likelihood of the data as

$$\ell_n(\theta) = \sum_{i=1}^n \ln p_\theta(\mathbf{y}_i | \mathbf{x}_i). \tag{1.2}$$

We will omit the subscript $n$ when it is clear from the context.

The corresponding random conditional log-likelihood $L_n(\theta)$ is obtained by replacing $(\mathbf{x}_i, \mathbf{y}_i)$ by the random variables $(\mathbf{X}_i, \mathbf{Y}_i)$.

**Theorem 1.1.** Let $(\mathbf{X}_1, \mathbf{Y}_1), \ldots, (\mathbf{X}_n, \mathbf{Y}_n)$ be i.i.d. samples drawn according to $p_{\text{data}}(\mathbf{x}, \mathbf{y})$. If the conditional cross entropy $H\big(p_{\text{data}}(\mathbf{y}|\mathbf{x}), p_\theta(\mathbf{y}|\mathbf{x}) | p_{\text{data}}(\mathbf{x})\big)$ is finite, then

$$\frac{1}{n} L_n(\theta) = \frac{1}{n} \sum_{i=1}^n \ln p_\theta(\mathbf{Y}_i | \mathbf{X}_i) \to -H\big(p_{\text{data}}(\mathbf{y}|\mathbf{x}), p_\theta(\mathbf{y}|\mathbf{x}) | p_{\text{data}}(\mathbf{x})\big) \text{ in probability.} \tag{1.3}$$

**Proof.** The theorem is a direct consequence of the weak law of large numbers.

This result provides a statistical justification for maximizing the empirical conditional log-likelihood. As the sample size increases, the empirical conditional log-likelihood converges to its expectation, whose maximizer $p_{\theta^*}$ corresponds precisely to the model conditional distribution that minimizes the conditional relative entropy with respect to the data-generating distribution $p_{\text{data}}$. When the data-generating distribution belongs to the probability model, maximizing the conditional log-likelihood is the conditional counterpart of maximum likelihood estimation. In this case, under certain regularity conditions, the empirical maximizers are consistent, that is, their corresponding conditional distributions converge to the true data-generating conditional distribution.

**Remarks.**

1. The negative conditional log-likelihood per sample,

$$-\frac{1}{n}\ell_n(\theta) = -\frac{1}{n}\sum_{i=1}^{n} \ln p_\theta(\mathbf{y}_i|\mathbf{x}_i),$$

   is also called the *empirical logarithmic loss (log loss)*. Its expectation with respect to $p_{\text{data}}$ is the conditional cross entropy. Thus, maximizing the conditional log-likelihood is equivalent to minimizing the empirical log loss. The optimal predictor under expected log loss, and the resulting operational interpretation of $H(X|Y)$ and $I(X;Y)$, are developed in Problem 1.1.

2. Supervised learning is one of several learning paradigms. We will also encounter *unsupervised learning* in this chapter, where we are given a dataset $\{\mathbf{x}_i\}_{i=1}^{n}$ without associated labels and seek to select a probability distribution $p_\theta(\mathbf{x})$ from a prescribed model that is closest to the unknown data-generating distribution $p_{\text{data}}$. As in supervised learning, relative entropy is a commonly used measure of divergence, although, as we will see, other objective functions are also used.

In the following, we demonstrate supervised learning through the classical models of linear regression and logistic regression.

### 1.1.1 Linear regression

In regression analysis, we specify a conditional pdf $f(y|\mathbf{x})$ and aim to predict the output (response) variable $Y$ given the input vector $\mathbf{x} = (x_1, x_2, \dots, x_K)$. The predictor $r(\mathbf{x}) = \mathsf{E}(Y|\mathbf{X} = \mathbf{x})$ is called the *regression function*.

The regression function is estimated from a dataset of input–output pairs $\{(\mathbf{x}_i, y_i)\}_{i=1}^{n}$. However, this estimation can be challenging and typically requires imposing structural assumptions on the conditional pdf of $Y$ given the inputs. A particularly tractable approach assumes a Gaussian conditional pdf and an affine conditional mean. This naturally leads to the linear regression model.

**Definition 1.2.** The *linear regression probability model* consists of all conditional pdfs of the form

$$f_\theta(y|\mathbf{x}) = \mathrm{N}\Big(y; w_0 + \sum_{k=1}^{K} w_k x_k, \sigma^2\Big), \tag{1.4}$$

where $w_0, w_1, \dots, w_K \in \mathbb{R}$ are referred to as the regression coefficients (or weights), $\sigma^2$ is the model noise variance, and the parameter vector is $\theta = (w_0, w_1, \dots, w_K, \sigma^2)$.

Under this model, the regression function $r(\mathbf{x}) = \mathsf{E}(Y|\mathbf{X} = \mathbf{x}) = w_0 + \sum_{k=1}^{K} w_k x_k$ is affine in $\mathbf{x}$. Defining $\mathbf{w}^T = [w_0\ w_1\ \cdots\ w_K]$ and the augmented input vector $\tilde{\mathbf{x}} = [1\ x_1\ \cdots\ x_K]^T$,

we can rewrite the parameters more compactly as $\theta = (\mathbf{w}, \sigma^2)$ and the regression function as $r(\mathbf{x}) = \mathbf{w}^T\tilde{\mathbf{x}}$.

To estimate (or learn) the model parameters $\theta$, we use supervised learning, that is, we find the parameters $(\hat{\mathbf{w}}, \hat{\sigma}^2)$ that maximize the conditional log-likelihood of the data,

$$\ell(\mathbf{w}, \sigma^2) = \sum_{i=1}^{n} \ln \mathrm{N}(y_i; \mathbf{w}^T\tilde{\mathbf{x}}_i, \sigma^2) \tag{1.5}$$

$$= -\frac{n}{2}\ln(2\pi\sigma^2) - \frac{1}{2\sigma^2}\sum_{i=1}^{n}(y_i - \mathbf{w}^T\tilde{\mathbf{x}}_i)^2. \tag{1.6}$$

This optimization problem has a closed-form solution.

**Theorem 1.2.** The parameter estimates that maximize (1.6) are given by the familiar least squares formulas

$$\hat{\mathbf{w}} = (X^TX)^{-1}X^T\mathbf{y}, \qquad \hat{\sigma}^2 = \frac{1}{n}\sum_{i=1}^{n}(y_i - \hat{\mathbf{w}}^T\tilde{\mathbf{x}}_i)^2, \tag{1.7}$$

where

$$X = \begin{bmatrix} 1 & x_{11} & \cdots & x_{K1} \\ 1 & x_{12} & \cdots & x_{K2} \\ \vdots & \vdots & \vdots & \vdots \\ 1 & x_{1n} & \cdots & x_{Kn} \end{bmatrix}, \qquad \mathbf{y} = \begin{bmatrix} y_1 \\ y_2 \\ \vdots \\ y_n \end{bmatrix}, \tag{1.8}$$

and we assume that $X^TX$ is nonsingular and $\hat{\sigma}^2 > 0$.

The quantity $n\hat{\sigma}^2$ is referred to as the *residual sum of squares* (RSS).

**Prediction.** Once the parameters are learned, the estimated regression function $\hat{r}(\mathbf{x})$ can be used to predict the output corresponding to a previously unseen input $\mathbf{x}'$ via $\hat{y}' = \hat{\mathbf{w}}^T\tilde{\mathbf{x}}'$. The prediction performance is measured by the expected squared prediction error $\mathsf{E}\left[(Y - \hat{\mathbf{W}}^T\tilde{\mathbf{X}})^2\right]$, where the expectation is taken over both the training data (through $\hat{\mathbf{W}}$) and an independent pair $(\mathbf{X}, Y)$ drawn from the data-generating distribution. The next chapter develops fundamental limits on the achievable performance of such learning algorithms.

**Estimation.** Linear regression can also be viewed as a parameter estimation problem. In this setting, the data are assumed to be generated according to a linear model $Y = \mathbf{w}^T\tilde{\mathbf{X}} + Z$, where $Z \sim \mathrm{N}(0, \sigma^2)$ is independent of $\mathbf{X}$, with unknown parameters. The objective is to estimate the unknown "true" parameters $(\mathbf{w}^*, \sigma^{*2})$ from the training data. The performance of the estimator $\hat{\mathbf{W}}$ is then measured by its parameter estimation error, for example, its MSE $\mathsf{E}\left(\|\hat{\mathbf{W}} - \mathbf{w}^*\|_2^2\right)$, where the expectation is over the training data.

### 1.1.2 Binary logistic regression

In classification, we specify a conditional pmf $p(y|\mathbf{x})$, where $\mathbf{x} = (x_1, x_2, \ldots, x_K)$ is the input vector and $y$ is the class label, which takes values in a finite set $\mathcal{M}$ of size $M$. The goal is to predict the class $Y$ given the input vector. For example, the input vector may represent an email or features extracted from it, with class label $y \in \{\text{SPAM, NOT SPAM}\}$.

To perform classification, we use a dataset of input–output pairs $\{(\mathbf{x}_i, y_i)\}_{i=1}^n$ to estimate a model of the conditional pmf $p(y|\mathbf{x})$. We focus on logistic regression, a classical and widely used model for classification.

We first consider binary logistic regression ($\mathcal{M} = \{0, 1\}$). Define the *logistic function* (Figure 1.2) as

$$\sigma(a) = \frac{1}{1 + e^{-a}}, \quad a \in \mathbb{R}. \tag{1.9}$$

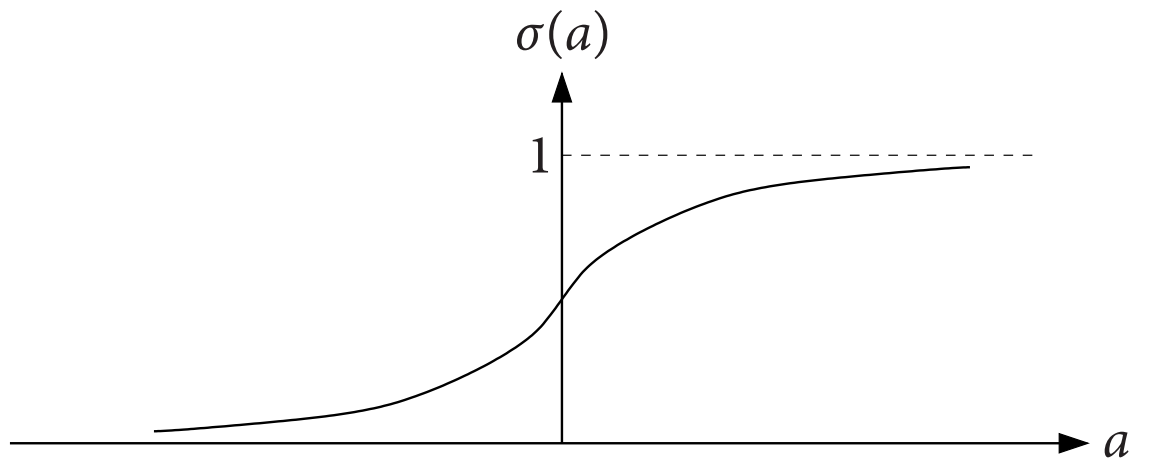

**Figure 1.2.** Logistic function.

**Definition 1.3.** The *logistic regression probability model* consists of all conditional pmfs of the form

$$p_\theta(y|\mathbf{x}) = \text{Bern}\Big(y; \sigma\big(w_0 + \sum_{j=1}^{K} w_j x_j\big)\Big) \tag{1.10}$$

as $\theta = \mathbf{w}^T = [w_0\ w_1\ \cdots\ w_K]$ ranges over $\mathbb{R}^{K+1}$.

As in linear regression, the augmented input vector is given by $\tilde{\mathbf{x}} = [1\ x_1\ \cdots\ x_K]^T$. We define the *logistic regression function* $p(\mathbf{x}) = \sigma(\mathbf{w}^T\tilde{\mathbf{x}})$, which gives the conditional probability $p_\theta(1|\mathbf{x})$. Given a labeled dataset $\{(\mathbf{x}_i, y_i)\}_{i=1}^n$, where $\mathbf{x}_i \in \mathbb{R}^K$ and $y_i \in \{0, 1\}$, we wish to learn the parameters $\mathbf{w}$. This is another instance of supervised learning, in which each input $\mathbf{x}_i$ is paired with a known label $y_i$: the parameters $\hat{\mathbf{w}}$ are determined by maximizing the conditional log-likelihood of the data,

$$\ell(\mathbf{w}) = \sum_{i=1}^{n} \ln \text{Bern}(y_i; p_i) \tag{1.11}$$

$$= \sum_{i=1}^{n} \big(y_i \ln p_i + (1 - y_i)\ln(1 - p_i)\big), \tag{1.12}$$

where $p_i = \sigma(\mathbf{w}^T\tilde{\mathbf{x}}_i)$, $i = 1, 2, \dots, n$. Hence, the negative log-likelihood of the data equals $n$ times the average cross entropy (or average log loss).

Unlike linear regression, even when a finite maximizer exists, there is no closed-form solution for $\hat{\mathbf{w}}$. However, the log-likelihood $\ell(\mathbf{w})$ is concave in $\mathbf{w}$ (Problem 1.2) and can be maximized numerically using, for example, *gradient ascent*—an iterative optimization algorithm of the form

$$\mathbf{w}^{(t+1)} = \mathbf{w}^{(t)} + \gamma \nabla \ell(\mathbf{w}^{(t)}), \ t = 1, 2, \ldots, \tag{1.13}$$

where $\gamma > 0$ is referred to as the *learning rate*. Using the observation that

$$\begin{aligned} \frac{d\sigma}{da} &= \frac{e^{-a}}{(1+e^{-a})^2} && (1.14) \\ &= \sigma(a)\frac{e^{-a}}{1+e^{-a}} && (1.15) \\ &= \sigma(a)(1-\sigma(a)), && (1.16) \end{aligned}$$

we can show that the gradient has the simple form

$$\nabla \ell(\mathbf{w}) = \sum_{i=1}^{n} (y_i - p_i)\tilde{\mathbf{x}}_i. \tag{1.17}$$

### 1.1.3 Multiclass logistic regression

Logistic regression can be extended to more than two classes. Let $\mathbf{a} \in \mathbb{R}^M$, and define the *softmax function* as

$$\text{softmax}(\mathbf{a}) = \left( \frac{e^{a_1}}{\sum_{m=1}^{M} e^{a_m}}, \ldots, \frac{e^{a_M}}{\sum_{m=1}^{M} e^{a_m}} \right). \tag{1.18}$$

Note that this naturally extends the logistic function to $M > 2$.

We denote the *categorical probability mass function* by $\text{Cat}(m; \mathbf{p})$, where $\mathbf{p} = (p_1, \ldots, p_M)$ with

$$p_m = \text{softmax}_m(\mathbf{a}) = \frac{e^{a_m}}{\sum_{m'=1}^{M} e^{a_{m'}}}, \quad m = 1, \ldots, M. \tag{1.19}$$

**Definition 1.4.** The *multiclass logistic regression probability model* consists of conditional pmfs of the form

$$p_\theta(y|\mathbf{x}) = \text{Cat}\big(y; \text{softmax}(\mathbf{w}_1^T\tilde{\mathbf{x}}, \ldots, \mathbf{w}_M^T\tilde{\mathbf{x}})\big), \tag{1.20}$$

where $\tilde{\mathbf{x}}^T = [1 \ x_1 \ \cdots \ x_K]$ is the augmented input vector, $\theta = [\mathbf{w}_1 \ \mathbf{w}_2 \ \cdots \ \mathbf{w}_M]^T$ is the weight matrix, and $y \in \{1, 2, \ldots, M\}$ is the output (class label).

As before, we are given a labeled dataset $\{(\mathbf{x}_i, y_i)\}_{i=1}^n$ and wish to learn the parameters $\hat{\mathbf{w}}_1, \ldots, \hat{\mathbf{w}}_M$ that maximize the conditional log-likelihood

$$\ell(\mathbf{w}_1, \ldots, \mathbf{w}_M) = \sum_{i=1}^{n} \sum_{m=1}^{M} y_{im} \ln p_{im}, \tag{1.21}$$

for $m = 1, \dots, M,\ i = 1, \dots, n,$

$$p_{im} = \mathrm{softmax}_m(\mathbf{w}_1^T \tilde{\mathbf{x}}_i, \dots, \mathbf{w}_M^T \tilde{\mathbf{x}}_i) \text{ and} \tag{1.22}$$

$$y_{im} = \begin{cases} 1 & \text{if } y_i = m, \\ 0 & \text{otherwise.} \end{cases} \tag{1.23}$$

This optimization problem again involves maximizing a concave function and its solution can be found using gradient ascent.

## 1.2 NEURAL NETWORKS

First we introduce the single-layer neural network as a natural extension of the linear and logistic regression functions.

### 1.2.1 Single-layer networks

A single-layer neural network, illustrated in Figure 1.3, consists of a set of inputs $\mathbf{x}$ and a set of outputs $\mathbf{y}$ related by

$$y_m = \phi_m\Big(w_{m0} + \sum_{k=1}^{K} w_{mk} x_k\Big),\ m = 1, \dots, M, \tag{1.24}$$

where $w_{m0}, w_{m1}, \dots, w_{mK}$ are the *weights* and $\phi_m$ is the *activation function* associated with output $m = 1, \dots, M$.

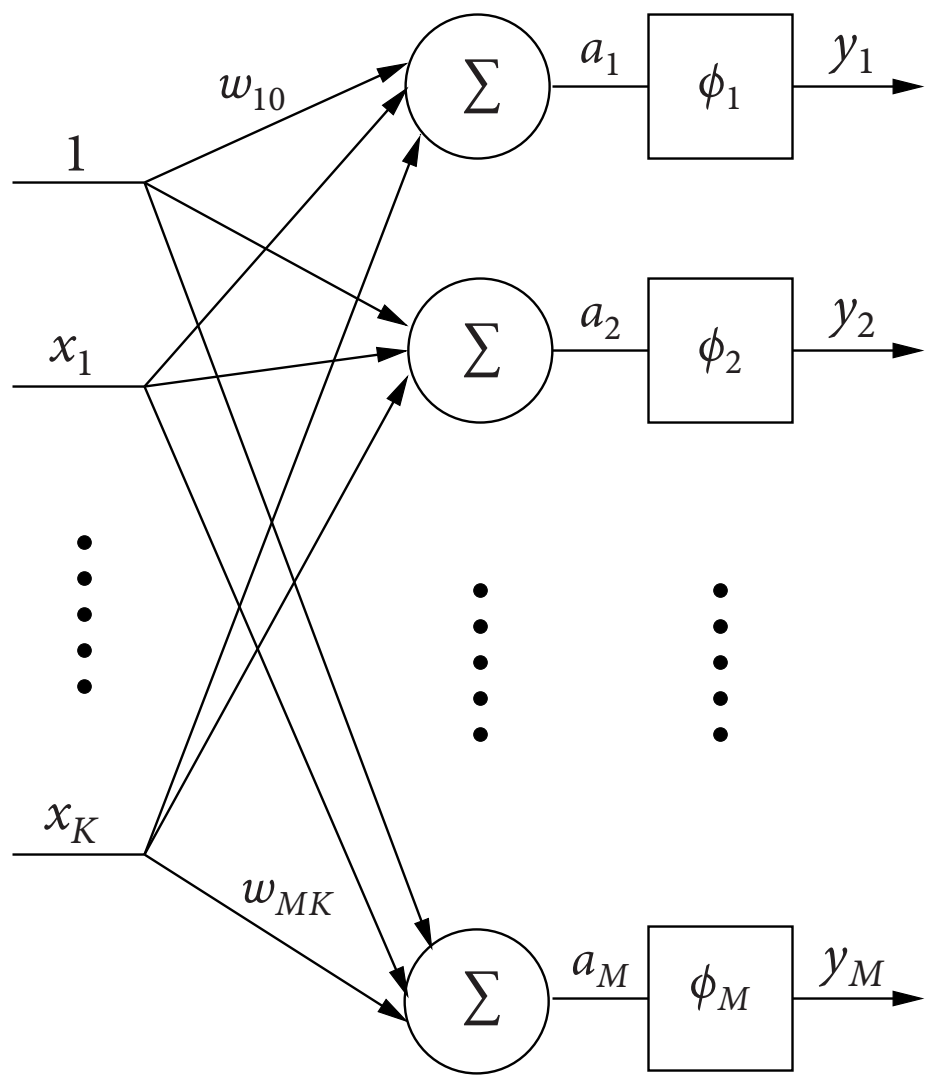

**Figure 1.3.** Single-layer neural network.

Examples of a single-layer neural network include:

- *Linear regression function.* Here $\phi_m$ is the identity function.
- *Logistic regression function.* Here $\phi_1(a) = \sigma(a)$.
- *Perceptron.* The earliest example of a 1-layer neural network is the *perceptron* in which the activation function is a Heaviside step function

$$\phi_m(a) = \begin{cases} 1 & \text{if } a \geq 0, \\ 0 & \text{if } a < 0. \end{cases}$$

- *Multiclass logistic regression.* Here $\phi_1, \ldots, \phi_M$ are replaced with a vector-valued function $\boldsymbol{\phi}(\mathbf{a}) = \text{softmax}(\mathbf{a})$.

### 1.2.2 Two-layer networks

The fact that the output of a single-layer neural network is a function of a weighted sum of the inputs limits its ability to approximate functions. One way to increase its expressive power is to add an additional layer of fixed, nonlinear *basis functions* such that

$$y_m = \phi_m\Big(w_{m0} + \sum_{k=1}^{K} w_{mk}\psi_k(\mathbf{x})\Big), \; m = 1, \ldots, M. \tag{1.25}$$

For example, assuming a single input $x$, the basis functions might be polynomial $\psi_k(x) = x^k$, "Gaussian" $\psi_k(x) = \exp\Big(-\frac{(x-\mu_k)^2}{2s^2}\Big)$, or logistic sigmoid $\psi_k(x) = \sigma\Big(\frac{x-\mu_k}{s}\Big)$, among others.

By increasing the number of basis functions, such models can approximate a wide class of functions arbitrarily well. Moreover, since the model remains linear in the weights, the derivations for linear and logistic regression in Sections 1.1.1 and 1.1.2 carry over naturally by replacing each input variable $x_k$ with $\psi_k(\mathbf{x})$. However, the effectiveness of this approach is limited in practice by the fact that the basis functions are fixed and cannot adapt to the training data.

This limitation can be overcome by learning the basis functions from the data rather than fixing them a priori. The resulting architecture consists of a *hidden layer* and an *output layer*. Following standard machine learning terminology, the input layer is not counted as a network layer; thus, this architecture is referred to as a two-layer neural network.

In the first layer, we form a linear combination of the inputs to obtain

$$a_m^{(1)} = w_{m0}^{(1)} + \sum_{k=1}^{K} w_{mk}^{(1)} x_k, \; m = 1, \ldots, M_1. \tag{1.26}$$

We then transform each $a_m^{(1)}$ using a nonlinear *activation function* to obtain the outputs of the first layer

$$z_m^{(1)} = \phi^{(1)}(a_m^{(1)}), \; m = 1, \ldots, M_1. \tag{1.27}$$

In the second layer, we linearly combine the $z_k^{(1)}$ values to obtain

$$a_m^{(2)} = w_{m0}^{(2)} + \sum_{k=1}^{M_1} w_{mk}^{(2)} z_k^{(1)}, \; m = 1, \ldots, M. \tag{1.28}$$

Finally, we transform each $a_m^{(2)}$ using another activation function to obtain the network outputs

$$y_m = \phi_m^{(2)}(a_m^{(2)}), \; m = 1, \ldots, M. \tag{1.29}$$

Examples of the activation functions used in the hidden layer include the logistic sigmoid $\sigma(a)$ (Figure 1.2), the Heaviside step function $\phi_1(a)$ (Figure 1.4(a)), and the *rectified linear unit* (ReLU) $\phi_2(a) = \max(0, a)$ (Figure 1.4(b)). The same activation function may be used in the output layer, although different choices are often made depending on the application. For example, in regression, the identity function is typically used, whereas in multiclass classification, the softmax function is used.

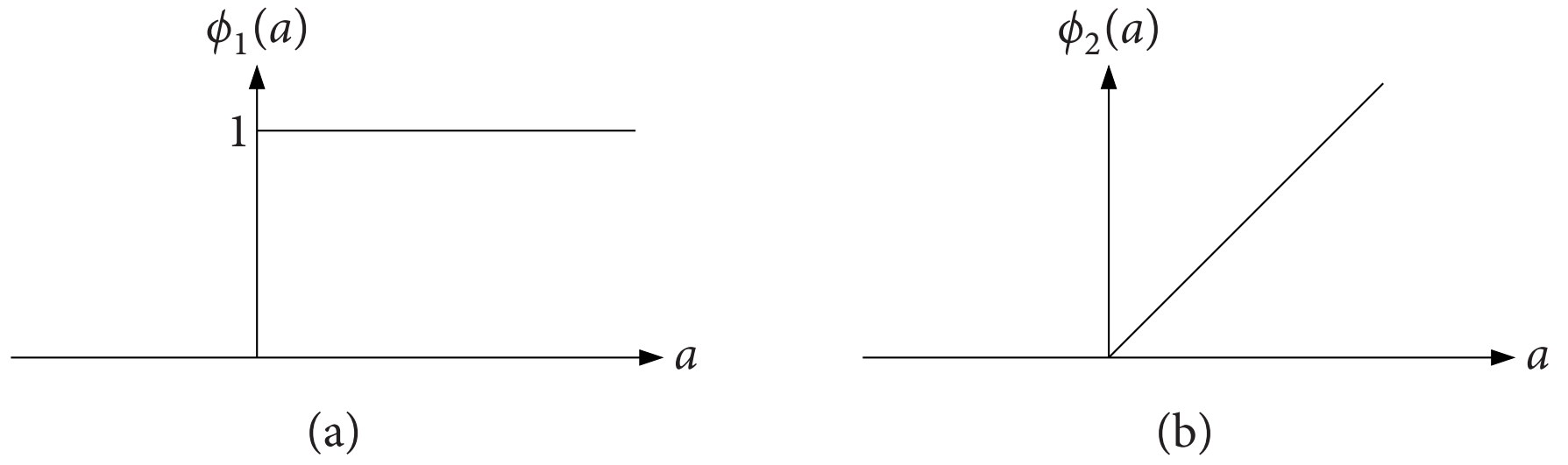

**Figure 1.4.** (a) Heaviside and (b) ReLU activation functions.

**Universal approximator.** The two-layer neural network described above,

$$y = \sum_{m=1}^{M} w_m^{(2)} \phi\Big(w_{m0}^{(1)} + \sum_{k=1}^{K} w_{mk}^{(1)} x_k\Big) \tag{1.30}$$

is a *universal function approximator*, in the sense that, with sufficiently many hidden units, it can uniformly approximate any continuous function on a compact set arbitrarily well. The following theorem, due to Leshno et al. (1993), characterizes the conditions under which this holds.

**Theorem 1.3.** Let $\phi : \mathbb{R} \to \mathbb{R}$ be a locally bounded and piecewise continuous function, where locally bounded means bounded on every bounded interval. Then the two-layer neural network in (1.30) is a universal approximator if and only if $\phi$ is not equal to a polynomial almost everywhere.

Both the logistic sigmoid and ReLU activation functions satisfy the conditions of the theorem and are nonpolynomial. Thus, two-layer neural networks using either activation function are universal function approximators.

Although the theorem establishes the expressive power of such networks, it only guarantees the existence of a network that can approximate any given continuous function on a compact set arbitrarily well. In particular, it does not provide guidance on the network size required to achieve a desired approximation accuracy, nor does it address how such networks can be efficiently trained in practice.

### 1.2.3 Deep networks

Some of the limitations of two-layer networks can be addressed by adding more hidden layers, which often reduce the number of parameters required to represent complex functions (Mhaskar, Liao, and Poggio 2017). Although such *deep neural networks* were once computationally impractical, advances in computing hardware and optimization algorithms, together with the availability of large datasets, have made it feasible to train networks with dozens or even hundreds of layers and billions of parameters.

We extend the description of the two-layer network to $L > 2$ layers as follows. Let the input vector be $\mathbf{z}^{(0)} = \tilde{\mathbf{x}}$, where $\tilde{\mathbf{x}} = [1\ x_1\ \cdots\ x_K]^T$. For each layer $l = 1, 2, \ldots, L-1$, let the output vector $\mathbf{z}^{(l)}$ have length $M_l + 1$, and for layer $L$, let the output vector $\mathbf{y}$ have length $M$. Define the weight matrices and their corresponding outputs as follows

$$W^{(l)} = \begin{bmatrix} (\mathbf{w}_1^{(l)})^T \\ (\mathbf{w}_2^{(l)})^T \\ \vdots \\ (\mathbf{w}_{M_l}^{(l)})^T \end{bmatrix}, \quad \mathbf{z}^{(l)} = \begin{bmatrix} 1 \\ z_1^{(l)} \\ \vdots \\ z_{M_l}^{(l)} \end{bmatrix}, \ l = 1, \ldots, L-1. \tag{1.31}$$

Then the output of layer $l = 1, \ldots, L-1$ can be expressed as

$$\mathbf{z}^{(l)} = \begin{bmatrix} 1 \\ \boldsymbol{\phi}^{(l)}\left(W^{(l)}\mathbf{z}^{(l-1)}\right) \end{bmatrix}, \tag{1.32}$$

where $\boldsymbol{\phi}^{(l)}(\mathbf{a}) = \left[\phi^{(l)}(a_1)\ \cdots\ \phi^{(l)}(a_{M_l})\right]^T$. The output of the network is

$$\mathbf{y} = \boldsymbol{\phi}^{(L)}\Big(W^{(L)}\mathbf{z}^{(L-1)}\Big), \tag{1.33}$$

where $\boldsymbol{\phi}^{(L)}$ may be componentwise, or may be a joint vector-valued function such as softmax. This deep neural network is illustrated in Figure 1.5.

### 1.2.4 Neural network-based regression and classification

A neural network (NN) can be used in regression and classification as follows.

**Regression.** The probability model with a neural regression function is of the form

$$p_\theta(y|\mathbf{x}) = \mathrm{N}(y; r_{\mathrm{NN}}(\mathbf{x}), \sigma^2), \tag{1.34}$$

where $r_{\mathrm{NN}}(\mathbf{x})$ is modeled by a neural network with the output layer having a single identity activation function unit. The number of layers, the number of units in each layer, and the activation functions used in the hidden layers are specified before training.

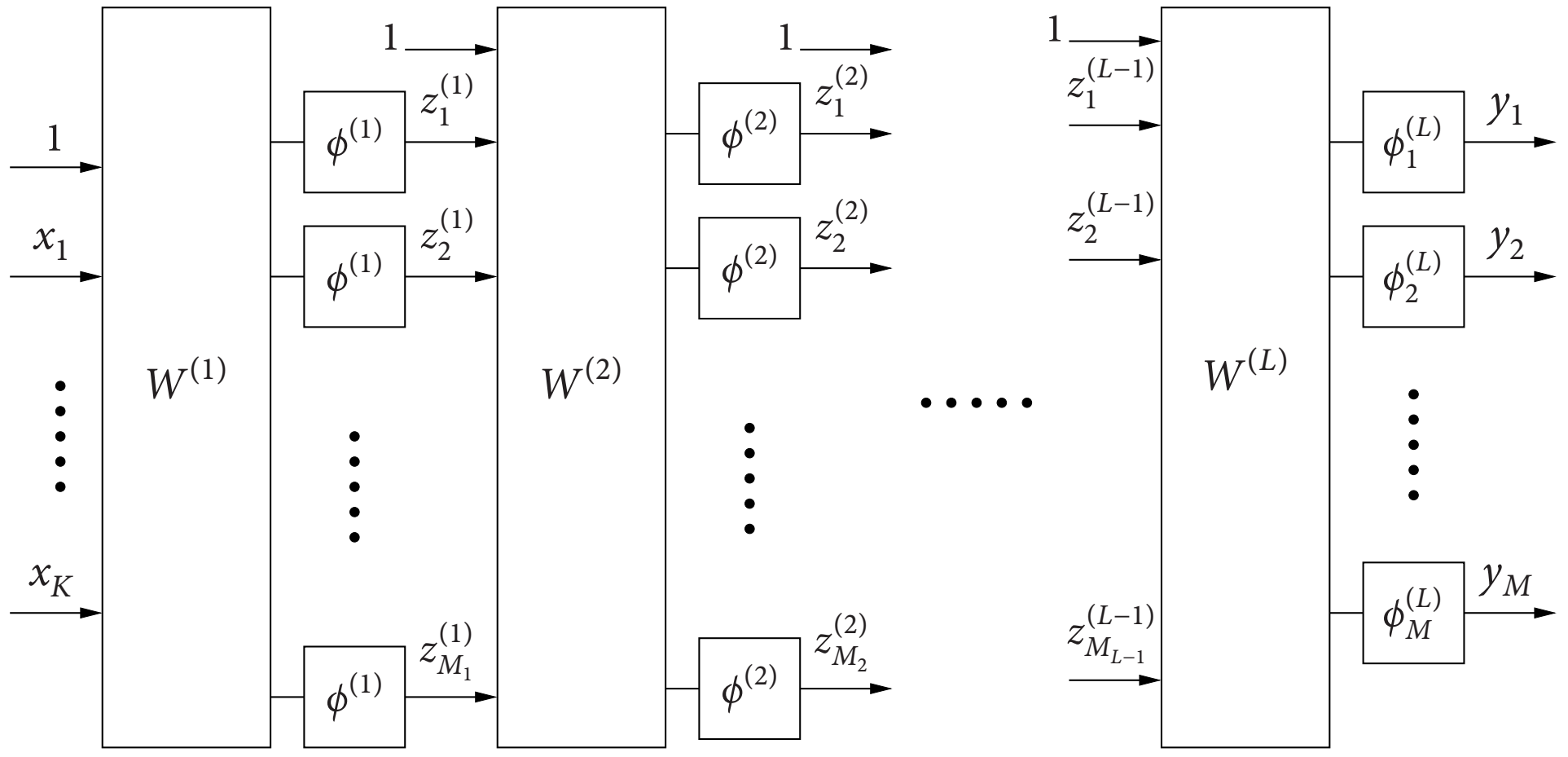

**Figure 1.5.** Deep feedforward neural network. Activation layers $1, \ldots, L-1$ are known as hidden layers.

**Two-class classification.** The probability model is of the form

$$p_\theta(y|\mathbf{x}) = \text{Bern}\big(y; p_{\text{NN}}(\mathbf{x})\big), \tag{1.35}$$

where $\theta = p_{\text{NN}}$ is modeled by a neural network whose output has a single logistic sigmoid activation unit.

**Multiclass classification.** The probability model is of the form

$$p_\theta(y|\mathbf{x}) = \text{Cat}\big(y; \mathbf{p}_{\text{NN}}(\mathbf{x})\big), \tag{1.36}$$

where $\theta = \mathbf{p}_{\text{NN}}$ is modeled by a neural network whose output layer uses the softmax activation function, i.e., $\mathbf{p}_{\text{NN}}(\mathbf{x}) = \text{softmax}(W^{(L)}\mathbf{z}^{(L-1)})$.

### 1.2.5 Remarks on deep neural networks

1. **Training.** As in linear and logistic regression, training a neural network is based on a dataset and seeks to find a set of weights that maximizes the log-likelihood, or more generally, optimizes a chosen objective function. Unlike these simpler models, the resulting optimization problem is typically highly non-convex, so there are no guarantees of finding a global optimum. In practice, neural networks employ activation functions that are differentiable almost everywhere, allowing gradient-based optimization methods to search for a local optimum, or more generally, a stationary point. The required gradients are computed efficiently by repeated application of the chain rule of differentiation, a procedure known as *backpropagation*. Training is further accelerated by stochastic gradient descent (SGD) and its variants, which estimate the gradient at each iteration using a small randomly selected subset of the data, called a *mini-batch*. Additional details and variations can be found in standard textbooks on deep learning.

2. **Network architectures.** The deep neural network architecture we described is a basic *feedforward* network. Other architectures, designed for specific applications, incorporate specialized components, such as *convolutional* layers, *self-attention* mechanisms, as used in *transformer-based* models, or feedback connections, as in *recurrent* neural networks. Further details and additional examples can be found in standard books on deep learning.

## 1.3 AUTOREGRESSIVE MODELS

We now turn to generative models, where the goal is to train a probability model and then use it to generate new samples that resemble the training dataset. For example, in natural language processing, the goal is to produce text reflecting the structure and statistical properties of a language; in music composition, to generate pieces consistent with a given genre; and in computer vision, to synthesize images or videos with desired content.

We first consider *autoregressive models*, which are particularly suited to data produced by *sequential* information sources, such as text, speech, and music.

**Definition 1.5.** An *autoregressive probability model* $\mathscr{P}$ consists of all pmfs (or pdfs) expressed via the chain rule for probability as

$$p_\theta(\mathbf{x}) = p_\theta(x_1)p_\theta(x_2|x_1)\dots p_\theta(x_K|x^{K-1}), \tag{1.37}$$

where $\mathbf{x} \in \mathcal{X}^K$ for some finite set $\mathcal{X}$ (or $\mathbf{x} \in \mathbb{R}^K$ for pdfs) and $\theta$ takes values in a parameter set $\Theta$.

The ordering of the variables $x_k$ corresponds to the order in which the data are generated, for example, the order of words or *tokens* in a block of text.

### 1.3.1 Training autoregressive models

Given an autoregressive probability model $\mathscr{P}$ and a dataset $\{\mathbf{x}_i\}_{i=1}^n$ drawn i.i.d. according to an unknown pmf $p_{\text{data}}$, we wish to select the pmf $p_\theta \in \mathscr{P}$ that is closest to $p_{\text{data}}$.

First assume that $\mathbf{x} \in \mathcal{X}^K$ for some finite set $\mathcal{X}$. We can "train" the model using the given dataset simply by estimating each conditional probability $p_\theta(x_k|x^{k-1})$, $k = 1, 2, \dots, K$, via a frequency table. However, as $k$ increases, the size of the table required to store these estimates grows exponentially, which makes this approach impractical. We can address this by approximating the autoregressive model by a Markov process of limited context. This approach was used by Shannon (1948) to estimate the entropy of English text. However, the text generated by such a model bears little resemblance to natural language. This suggests that generating sentences that can pass as English text requires a model with long contexts.

Another approach to training an autoregressive model is to parametrize each conditional distribution. For example, assuming binary-valued data, i.e., $\mathbf{x} \in \{0, 1\}^K$, we could use a separate logistic regression model for each step $k$:

$$p_{\theta_k}(x_k | x^{k-1}) = \mathrm{Bern}\Big(x_k; \sigma\Big(w_{0k} + \sum_{j=1}^{k-1} w_{jk} x_j\Big)\Big), \tag{1.38}$$

and train each model using supervised learning. For large $K$, this approach requires training $K$ different models, one for each conditional distribution, which is inefficient and scales poorly. Using a more powerful model for each conditional, such as a dedicated neural network, would only worsen this scaling problem, making it computationally expensive. More importantly, this approach fails to capture the intuition that the underlying rules of a sequence (e.g., grammar in text) should be consistent across different positions.

A more effective and scalable solution is to use a single neural network with a shared set of parameters $\theta$ to model all the conditional distributions. The network takes as input a variable-length context $x^{k-1}$ and outputs the parameters for the distribution of the next token $x_k$ for every $k = 2, \ldots, K$. The model is trained jointly by maximizing the log-likelihood of the dataset $\{\mathbf{x}_i\}_{i=1}^n$:

$$\ell(\theta) = \sum_{i=1}^{n} \sum_{k=2}^{K} \ln p_\theta(x_{ki} | x_{1i}^{k-1}),$$

where $x_{1i}^{k-1} = (x_{1i}, x_{2i}, \ldots, x_{(k-1)i})$, and the distributions of the initial tokens $p(x_{1i})$ are assumed fixed. This is the principle underlying neural network-based autoregressive architectures such as transformers (Vaswani et al. 2017).

For discrete data like text, the network typically outputs the parameters of a softmax distribution over the vocabulary $\mathcal{X}$. For continuous data, the network can be configured to output the parameters of a more complex distribution, such as a Gaussian mixture model:

$$p_\theta(x_k | x^{k-1}) = \sum_{j=1}^{d} \pi_j(x^{k-1}) \cdot \mathrm{N}\Big(x_k; \mu_j(x^{k-1}), \sigma_j^2(x^{k-1})\Big), \tag{1.39}$$

where the mixture weights $\pi_j$, means $\mu_j$, and variances $\sigma_j^2$ are all outputs of the single neural network parameterized by $\theta$.

### 1.3.2 Remarks on autoregressive models

1. **Prediction.** A trained autoregressive model $p_{\hat\theta}(\mathbf{x})$ can be used for prediction. Given an observed prefix of a sequence, $a^{k-1}$, the model can predict the next element by evaluating the distribution $p_{\hat\theta}(x_k | a^{k-1})$.

2. **Sampling.** The predictive distribution of an autoregressive model provides a natural mechanism for sequence generation. A common approach is *ancestral sampling*: first draw $a_1$ from $p_{\hat\theta}(x_1)$; then draw $a_2$ from $p_{\hat\theta}(x_2 | a_1)$; and continue sequentially until a full sequence $a^K$ is obtained. This is the basic mechanism by which autoregressive large language models generate text.

3. **Inference.** In artificial intelligence, prediction and generation of samples from a model are commonly referred to as *inference*. This usage should not be confused with the notion of statistical inference.

4. **Perplexity.** In natural language processing—particularly for large language models (LLMs) (OpenAI 2023)—performance is typically evaluated on a test dataset using the empirical *perplexity*, defined as the exponential of the per-token negative log-likelihood (see Problem 1.3):

$$\text{Perplexity}(\theta) = \left( \prod_{i=1}^{n} \prod_{k=2}^{K} \frac{1}{p_\theta(x_{ki}|x_{1i}^{k-1})} \right)^{1/(n(K-1))}. \tag{1.40}$$

A lower perplexity indicates a better model fit to the test data, as it corresponds to lower average per-token negative log-likelihood.

5. **Scaling laws.** Empirical studies of large language models have found that the test log loss (cross entropy), i.e., the logarithm of the perplexity, decreases approximately according to power laws as the number of model parameters, the size of the training dataset, and the computational resources used in training increase (Kaplan et al. 2020). These scaling laws help predict the performance of larger models from experiments with smaller ones. They also provide guidance on how to allocate a fixed training compute budget between model size and training data. In particular, Hoffmann et al. (2022) found that compute-optimal training requires increasing the number of model parameters and training tokens in approximately equal proportion.

6. **Entropy rate of English.** An autoregressive model serves as a sequential probability estimator: given the context $x^{k-1}$, it assigns a conditional probability $p_{\hat\theta}(x_k|x^{k-1})$ to the next token. The cross entropy rate between the data process $p_{\text{data}}(x^k)$ and the learned model process $p_{\hat\theta}(x^k)$, $k = 1, 2, \ldots$, is defined by

$$\overline{H}(p_{\text{data}}, p_{\hat\theta}) = \lim_{k\to\infty} \frac{1}{k} \mathsf{E}_{p_{\text{data}}} \left( \log \frac{1}{p_{\hat\theta}(X^k)} \right),$$

provided the limit exists. Since, for every $k \geq 1$,

$$\mathsf{E}_{p_{\text{data}}} \left( \log \frac{1}{p_{\hat\theta}(X^k)} \right) = H(p_{\text{data}}(x^k)) + D(p_{\text{data}}(x^k) \| p_{\hat\theta}(x^k)) \geq H(p_{\text{data}}(x^k)),$$

the cross entropy rate is an upper bound on the entropy rate of the data process, with smaller relative entropy rates yielding tighter bounds.

Given $m$ independent test samples $\mathbf{x}_i$ drawn from $p_{\text{data}}(\mathbf{x})$, the cross entropy rate can be estimated (in bits per token) by

$$\frac{1}{mK} \sum_{i=1}^{m} \sum_{k=1}^{K} \log \frac{1}{p_{\hat\theta}(x_{ki}|x_{1i}^{k-1})}.$$

For sufficiently large $K$ and $m$, this yields an approximate upper bound on the entropy rate of English, as discussed earlier.

7. **Sequential probability estimation and compression.** The sequential probability estimates produced by an autoregressive model can be used for data compression. If the model $p_{\hat{\theta}}$ is available to both the encoder and decoder, arithmetic coding can encode $x_2^n$ given $x_1$ using approximately $-\sum_{k=2}^{n} \log p_{\hat{\theta}}(x_k|x^{k-1})$ bits.
   For sources such as English text, neural sequential probability estimators can achieve substantially better compression rates than classical sequential estimation algorithms, albeit at significantly higher computational complexity and with the need for offline training; see, for example, (Delétang et al. 2024). These compression-rate improvements, however, assume that the trained model is already shared between the encoder and decoder. If the model must also be transmitted, its description length should be included in the total coding length.

8. **Lempel–Ziv coding as an autoregressive learning algorithm.** The LZ78 compression algorithm may be viewed as a form of online autoregressive learning. Rather than fitting a parameterized conditional probability model to a separate training dataset, LZ78 sequentially learns recurring contexts and their continuations from the sequence being processed. In fact, the LZ78 parsing can be used to construct a sequential probability assignment for the next symbol given the preceding symbols, whose normalized log loss asymptotically agrees with the normalized LZ78 code length (Sagan and Weissman 2026). Thus, LZ78 provides a nonparametric analogue of autoregressive learning that operates online, without a separate training phase.

## 1.4 LATENT VARIABLE MODELS AND THE EVIDENCE LOWER BOUND

In many applications, we are only able to collect data from a subset of the variables. The unobserved (so-called *latent*) variables may be, for example, a subset of the pixel values of an image, high-level *features* such as facial characteristics that provide a more compact description of the image, or a clustering of the data into groups sharing common characteristics.

**Definition 1.6.** A *latent variable probability model* consists of joint distributions $p_\theta(\mathbf{z}, \mathbf{x})$, where the observed variable is $\mathbf{x} \in \mathbb{R}^M$, and the latent variable is $\mathbf{z} \in \mathbb{R}^K$.

We are given a dataset $\{\mathbf{x}_i\}_{i=1}^n$ and wish to find the closest marginal of $\mathbf{X}$, $p_\theta(\mathbf{x})$, to $p_{\text{data}}(\mathbf{x})$. We can try to model $p_\theta(\mathbf{x})$ directly, but if we knew $\mathbf{z}$, modeling $p_\theta(\mathbf{x}|\mathbf{z})$ would presumably be simpler. We are also often interested in the posterior distribution of $\mathbf{Z}$ given $\mathbf{X}$, $p_\theta(\mathbf{z}|\mathbf{x})$.

Since latent variable models are trained solely using samples from the marginal distribution $p_{\text{data}}(\mathbf{x})$ rather than using paired samples from the joint distribution $p_{\text{data}}(\mathbf{z}, \mathbf{x})$, as in supervised learning, they fall under the setting of *unsupervised learning*. As before, the goal is to select a probability distribution $p_\theta(\mathbf{x})$ from a given model that is closest to the unknown data-generating distribution $p_{\text{data}}$ using a divergence measure such as relative entropy.

We begin with a classical latent variable model: probabilistic principal component analysis (PPCA). For PPCA, the maximum-likelihood estimates of the model parameters admit closed-form expressions. For the more complex models considered in Sections 1.5 and 1.6, however, evaluating the log-likelihood $\ell(\theta) = \ln \int p_\theta(\mathbf{x}|\mathbf{z})p(\mathbf{z})\, d\mathbf{z}$ is computationally intractable. This motivates the introduction of the evidence lower bound (ELBO), a tractable lower bound on the log-likelihood that serves as a surrogate objective for training.

### 1.4.1 Probabilistic principal component analysis

Principal component analysis (PCA) is a classical statistical analysis technique for identifying low-dimensional structure in high-dimensional datasets. PCA projects data onto a lower-dimensional subspace that maximizes the variance of the projected data, with the goal of retaining its most important features. By disregarding directions of low variance, PCA can also reduce noise when it is concentrated in these directions.

PCA can also be formulated as a probabilistic latent variable model.

**Definition 1.7.** The *Probabilistic Principal Component Analysis (PPCA)* is a latent variable model in which $(\mathbf{Z}, \mathbf{X})$ are jointly Gaussian and satisfy

$$f(\mathbf{z}) = \mathrm{N}(\mathbf{z}; \mathbf{0}, I_K), \tag{1.41}$$

$$f_\theta(\mathbf{x}|\mathbf{z}) = \mathrm{N}\big(\mathbf{x}; W\mathbf{z} + \boldsymbol{\mu}, \sigma^2 I_M\big). \tag{1.42}$$

Here the observed variable $\mathbf{x} \in \mathbb{R}^M$, the latent variable $\mathbf{z} \in \mathbb{R}^K$, where typically $K \ll M$, $W$ is an $M \times K$ matrix, and $\boldsymbol{\mu}$ is an $M$-dimensional vector.

The parameter $\theta = (W, \boldsymbol{\mu}, \sigma^2)$ is to be learned from a dataset $\{\mathbf{x}_i\}_{i=1}^n$. Since $f(\mathbf{z})$ and $f_\theta(\mathbf{x}|\mathbf{z})$ are Gaussian, $f_\theta(\mathbf{x})$ and $f_\theta(\mathbf{z}|\mathbf{x})$ are also Gaussian and are given by

$$f_\theta(\mathbf{x}) = \mathrm{N}(\mathbf{x}; \boldsymbol{\mu}, C), \tag{1.43}$$

$$f_\theta(\mathbf{z}|\mathbf{x}) = \mathrm{N}\big(\mathbf{z}; A^{-1}W^T(\mathbf{x} - \boldsymbol{\mu}), \sigma^2 A^{-1}\big), \tag{1.44}$$

where $C = WW^T + \sigma^2 I$ and $A = W^T W + \sigma^2 I$.

**Training.** As before, the parameters are estimated by maximizing the log-likelihood of the data,

$$\ell(\theta) = \sum_{i=1}^n \ln f_\theta(\mathbf{x}_i) \tag{1.45}$$

$$= -\frac{nM}{2} \ln\left(2\pi\left|WW^T + \sigma^2 I\right|^{1/M}\right) - \frac{1}{2}\sum_{i=1}^n (\mathbf{x}_i - \boldsymbol{\mu})^T (WW^T + \sigma^2 I)^{-1}(\mathbf{x}_i - \boldsymbol{\mu}). \tag{1.46}$$

This optimization problem has a closed-form solution involving the same principal eigenvectors of the data covariance matrix as standard PCA.

**Theorem 1.4.** The solution to the optimization problem (1.46) is given by

$$\hat{\boldsymbol{\mu}} = \frac{1}{n}\sum_{i=1}^{n} \mathbf{x}_i, \tag{1.47}$$

$$\hat{\sigma}^2 = \frac{1}{M-K}\sum_{j=K+1}^{M} \lambda_j, \tag{1.48}$$

$$\hat{W} = U(\Lambda - \hat{\sigma}^2 I)^{1/2} V, \tag{1.49}$$

where $\Lambda$ is a $K \times K$ diagonal matrix whose diagonal elements correspond to the largest $K$ eigenvalues $\lambda_j$ of the data covariance matrix $S = (1/n)\sum_{i=1}^{n}(\mathbf{x}_i - \hat{\boldsymbol{\mu}})(\mathbf{x}_i - \hat{\boldsymbol{\mu}})^T$ arranged in decreasing order (and we assume that $\sum_{j=K+1}^{M} \lambda_j > 0$), $U$ is the $M \times K$ matrix whose columns are the $K$ eigenvectors corresponding to the $K$ largest eigenvalues of $S$, and $V$ is a $K \times K$ orthogonal matrix, i.e., $V^T V = V V^T = I$.

The orthogonal matrix $V$ reflects the fact that the latent representation is identifiable only up to an orthogonal transformation. A proof of this result is given in (Tipping and Bishop 1999).

### 1.4.2 Evidence Lower Bound (ELBO)

For PPCA, a simple latent variable model, the log-likelihood can be maximized in closed form. For more complex latent variable models, however, evaluating the log-likelihood is computationally intractable. We therefore introduce the evidence lower bound (ELBO), which provides a tractable surrogate objective for training such models.

We assume a probability model specified by joint pdfs $f_\theta(\mathbf{z}, \mathbf{x})$. The same derivation applies without modification to joint pmfs, and more generally, to models involving mixtures of discrete latent variables and continuous observations.

**Theorem 1.5** (ELBO)**.** For a data outcome $\mathbf{x}$ and any pdf $g(\mathbf{z})$,

$$\ln f_\theta(\mathbf{x}) \geq \mathsf{E}_{g(\mathbf{z})}\left(\ln f_\theta(\mathbf{x}|\mathbf{Z})\right) - D\big(g(\mathbf{z})\|f_\theta(\mathbf{z})\big) \tag{1.50}$$

$$= \mathsf{E}_{g(\mathbf{z})}\Big(\ln f_\theta(\mathbf{x}, \mathbf{Z}) - \ln g(\mathbf{Z})\Big). \tag{1.51}$$

Equality holds if and only if $g(\mathbf{z}) = f_\theta(\mathbf{z}|\mathbf{x})$ almost everywhere.

**Proof.** The proof is an immediate consequence of the variational characterization of relative entropy, which states that for any pdfs $g(\mathbf{z})$ and $f(\mathbf{z})$ and any function $\phi : \mathbb{R}^K \to \mathbb{R}$,

$$D(g\|f) \geq \mathsf{E}_g(\phi(\mathbf{Z})) - \ln \mathsf{E}_f\left(e^{\phi(\mathbf{Z})}\right). \tag{1.52}$$

For a fixed $\mathbf{x}$, we set $f(\mathbf{z}) = f_\theta(\mathbf{z})$ and $\phi(\mathbf{z}) = \ln f_\theta(\mathbf{x}|\mathbf{z})$. Then

$$\mathsf{E}_{f_\theta}\left(e^{\phi(\mathbf{Z})}\right) = \int f_\theta(\mathbf{x}|\mathbf{z}) f_\theta(\mathbf{z})\, d\mathbf{z} = f_\theta(\mathbf{x}),$$

which yields the desired inequality. The equality condition follows from that of the variational representation result, giving $g(\mathbf{z}) = f_\theta(\mathbf{z}|\mathbf{x})$ almost everywhere.

The gap between the log-likelihood and the ELBO is precisely the relative entropy $D(g(\mathbf{z}) \| f_\theta(\mathbf{z}|\mathbf{x}))$. Hence, the closer $g(\mathbf{z})$ is to the posterior distribution $f_\theta(\mathbf{z}|\mathbf{x})$, the tighter the lower bound. Since the auxiliary distribution $g(\mathbf{z})$ may depend on the observation $\mathbf{x}$, the ELBO is more generally written as

$$\ln f_\theta(\mathbf{x}) \geq \mathsf{E}_{g(\mathbf{z}|\mathbf{x})}\Big( \ln f_\theta(\mathbf{x}, \mathbf{Z}) - \ln g(\mathbf{Z}|\mathbf{x})\Big),$$

with equality if and only if $g(\mathbf{z}|\mathbf{x}) = f_\theta(\mathbf{z}|\mathbf{x})$ almost everywhere.

**Remark.** The ELBO comprises two complementary terms. The first encourages the model to reconstruct the observed data accurately through the conditional distribution $f_\theta(\mathbf{x}|\mathbf{z})$. The second encourages the approximate posterior $g(\mathbf{z}|\mathbf{x})$ to remain close to the prior $f_\theta(\mathbf{z})$. As a result, latent variables sampled directly from the prior have a distribution similar to the latent representations learned from the training data.

In the following two sections, we demonstrate how the ELBO is used to train neural network-based latent variable models. By maximizing the ELBO, which is typically much more tractable than direct log-likelihood maximization, we aim to obtain models that perform well in practice.

## 1.5 VARIATIONAL AUTOENCODERS

Autoencoders are latent variable models trained to reconstruct their inputs by learning efficient latent representations of the observed data. Source coding, where a source sequence is mapped to a compressed representation that is then decoded to recover the original sequence exactly or approximately, fits naturally into this framework. Unlike source coding, however, autoencoders are typically used not only for reconstruction but also for learning latent structure that supports the generation of new samples resembling the training data.

A variational autoencoder (VAE) is an autoencoder that employs flexible parametric probability models for both the encoder and decoder. PPCA can be viewed as a VAE with encoder $f(\mathbf{z}|\mathbf{x})$ in (1.44) and decoder $f(\mathbf{x}|\mathbf{z})$ in (1.42). However, its linear-Gaussian structure limits its ability to capture complex data distributions.

VAEs generalize PPCA by replacing these linear mappings with neural networks that parameterize the mean and covariance of the conditional Gaussian distribution.

**Definition 1.8.** The *variational autoencoder* is a latent variable model consisting of all pdfs of the form

$$f(\mathbf{z}) = \mathrm{N}(\mathbf{z}; \mathbf{0}, I), \tag{1.53}$$

$$f_\theta(\mathbf{x}|\mathbf{z}) = \mathrm{N}\big(\mathbf{x}; \boldsymbol{\mu}_\theta(\mathbf{z}), \Sigma_\theta(\mathbf{z})\big). \tag{1.54}$$

While in the above model $f_\theta(\mathbf{x}|\mathbf{z})$ is simple, the marginal $f_\theta(\mathbf{x})$ and posterior $f_\theta(\mathbf{z}|\mathbf{x})$ can be highly complex, providing great modeling flexibility but posing significant training challenges.

**Training a VAE.** The main challenge in training VAEs is that, because the conditional distributions are parameterized by neural networks, direct computation of the log-likelihood is impractical. Consequently, given a dataset $\{\mathbf{x}_i\}_{i=1}^n$, a VAE is trained by maximizing the evidence lower bound (ELBO) on the data log-likelihood (Theorem 1.5); that is, by maximizing

$$\mathcal{L}(\theta, \mathbf{g}) = \sum_{i=1}^{n} \big[ \mathsf{E}_{g(\mathbf{z}|\mathbf{x}_i)} \big( \ln f_\theta(\mathbf{x}_i|\mathbf{Z})\big) - D\big(g(\mathbf{z}|\mathbf{x}_i) \| f(\mathbf{z})\big)\big], \tag{1.55}$$

where $g(\mathbf{z}|\mathbf{x}_i)$ is an arbitrary variational distribution associated with data point $\mathbf{x}_i$. For fixed $\theta$, one could optimize a separate variational distribution $g_i(\mathbf{z})$ for each observation $\mathbf{x}_i$. These local optimizations are coupled through the subsequent optimization of the shared parameter $\theta$. In the variational autoencoder, this family of distributions is instead approximated by a single conditional distribution $g_\phi(\mathbf{z}|\mathbf{x})$, where $\phi$ is the parameter set of a neural network known as the *encoder*. This approach is referred to as *amortized inference*. A common choice of the encoder is a product Gaussian distribution of the form

$$g_\phi(\mathbf{z}|\mathbf{x}) = \prod_{k=1}^{K} \mathrm{N}\big(z_k; \mu_k(\mathbf{x}, \phi), \sigma_k^2(\mathbf{x}, \phi)\big). \tag{1.56}$$

With amortized inference, we would now jointly train two neural networks, the encoder $\phi$ and the decoder $\theta$, by maximizing the ELBO

$$\mathcal{L}(\theta, \phi) = \sum_{i=1}^{n} \Big[ \mathsf{E}_{g_\phi(\mathbf{z}|\mathbf{x}_i)} \big( \ln f_\theta(\mathbf{x}_i|\mathbf{Z})\big) - D(g_\phi(\mathbf{z}|\mathbf{x}_i) \| f(\mathbf{z}))\Big]. \tag{1.57}$$

Substituting from (1.56), the second term has a simple expression,

$$\begin{aligned}\mathcal{L}(\theta, \phi) = \sum_{i=1}^{n} \Big[ &\mathsf{E}_{g_\phi(\mathbf{z}|\mathbf{x}_i)} \big( \ln f_\theta(\mathbf{x}_i|\mathbf{Z})\big) \\ &+ \frac{1}{2}\sum_{k=1}^{K} \big(1 + \ln \sigma_k^2(\mathbf{x}_i, \phi) - \mu_k^2(\mathbf{x}_i, \phi) - \sigma_k^2(\mathbf{x}_i, \phi)\big)\Big].\end{aligned} \tag{1.58}$$

**Remark.** Computing the above expression remains challenging because the first term requires evaluating expectations. To overcome this difficulty, the expectation is approximated by a sample average over outcomes of $\mathbf{Z}$ generated from the variational posterior $g_\phi(\mathbf{z}|\mathbf{x})$. The parameters $\phi$ and $\theta$ are then updated iteratively until convergence (see (Bishop and Bishop 2024)).

To summarize, the variational autoencoder illustrated in Figure 1.6 consists of an encoder and a decoder. The encoder uses a neural network to model the parameters $\phi$ of a Gaussian approximation $g_\phi(\mathbf{z}|\mathbf{x})$ to the posterior, and the decoder employs another neural network to model the parameters $\theta$ of a Gaussian conditional pdf $f_\theta(\mathbf{x}|\mathbf{z})$. The two networks are simultaneously trained using stochastic gradient ascent to find the estimates $\hat{\phi}$ and $\hat{\theta}$. After training, the decoder can be used to generate a new data sample by first sampling a latent variable $\tilde{\mathbf{z}}$ from the prior $f(\mathbf{z}) = \mathrm{N}(\mathbf{z}; \mathbf{0}, I)$, and then passing it through the decoder to generate a sample $\tilde{\mathbf{x}}$. The encoder can be used to compute a latent representation $\mathbf{z}'$ of a new input data point $\mathbf{x}'$.

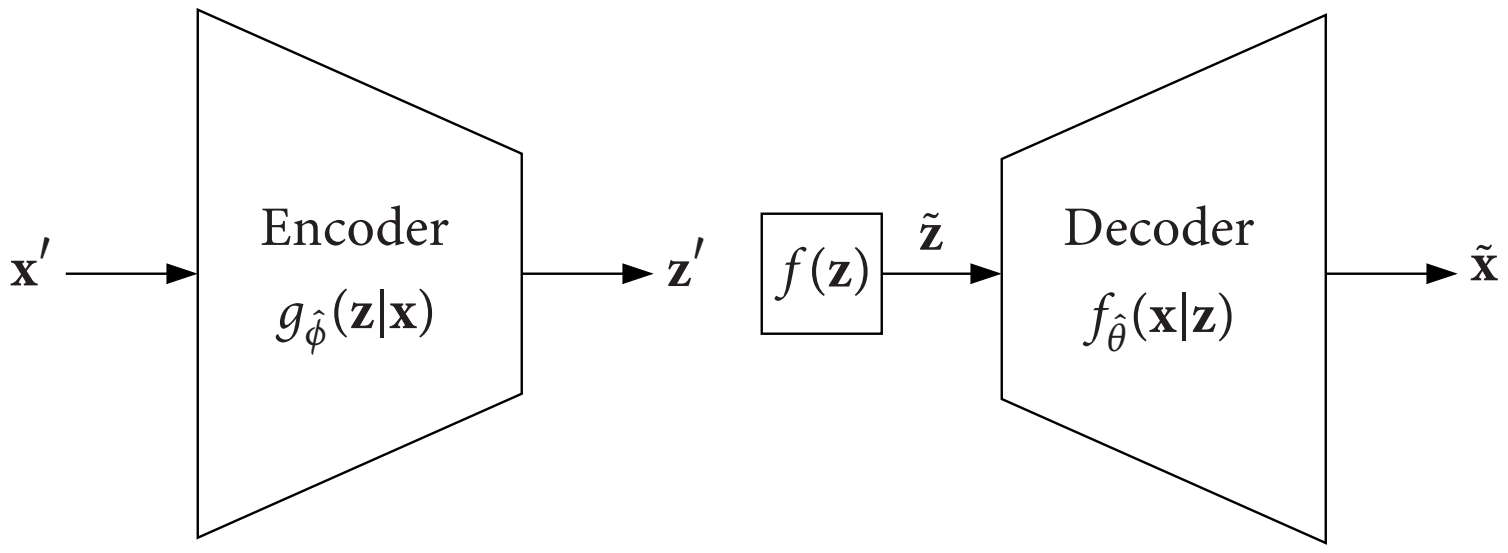

**Figure 1.6.** Variational autoencoder (VAE).

## 1.6 DIFFUSION MODELS

Generative diffusion models are inspired by the physical process of diffusion in which particles spread from regions of higher concentration to regions of lower concentration due to random motion, eventually reaching equilibrium. In physical systems, this process is associated with increasing thermodynamic entropy. Constructing a generative model requires explicitly learning a time-reversed version of these dynamics.

**Definition 1.9.** The *generative diffusion model* is a "hierarchical" latent variable model defined by a reverse-time (*backward*) Markov process specified by

$$f(\mathbf{z}_T) = \mathrm{N}(\mathbf{z}_T; \mathbf{0}, I), \tag{1.59}$$

$$f_\theta(\mathbf{z}_{t-1}|\mathbf{z}_t) = \mathrm{N}\big(\mathbf{z}_{t-1}; \boldsymbol{\mu}_t(\mathbf{z}_t), \beta'_t I\big), \quad t = 2, \ldots, T, \tag{1.60}$$

$$f_\theta(\mathbf{x}|\mathbf{z}_1) = \mathrm{N}(\mathbf{x}; \boldsymbol{\mu}_1(\mathbf{z}_1), \beta'_1 I), \tag{1.61}$$

where all the variables have the same dimension $K$, and $\beta'_t$, $t = 1, 2, \ldots, T$, are constants to be specified later. This model is parametrized by $\theta = \{\boldsymbol{\mu}_t(\mathbf{z}_t)\}_{t=1}^T$, which is modeled by a neural network.

We note that although the corresponding forward diffusion process is Markov, its conditional distributions $f_\theta(\mathbf{z}_t|\mathbf{z}_{t-1})$ are not, in general, Gaussian. We return to this point later.

### 1.6.1 Training diffusion models

Given a dataset $\{\mathbf{x}_i\}_{i=1}^n$, our goal is to learn a probability distribution $f_\theta(\mathbf{x})$ that approximates the data-generating distribution $f_{\text{data}}(\mathbf{x})$. Training is based on a forward process that gradually corrupts the data by adding noise until the resulting samples become indistinguishable from pure noise. The model is then trained to approximate the corresponding reverse process in (1.60) and (1.61), which progressively removes the noise and thereby recovers the data distribution. The forward process used during training and the learned reverse process used for sampling are illustrated in Figure 1.7. The particular noise schedule, number of steps ($T = 1000$), and training procedure used in this example are specific choices.

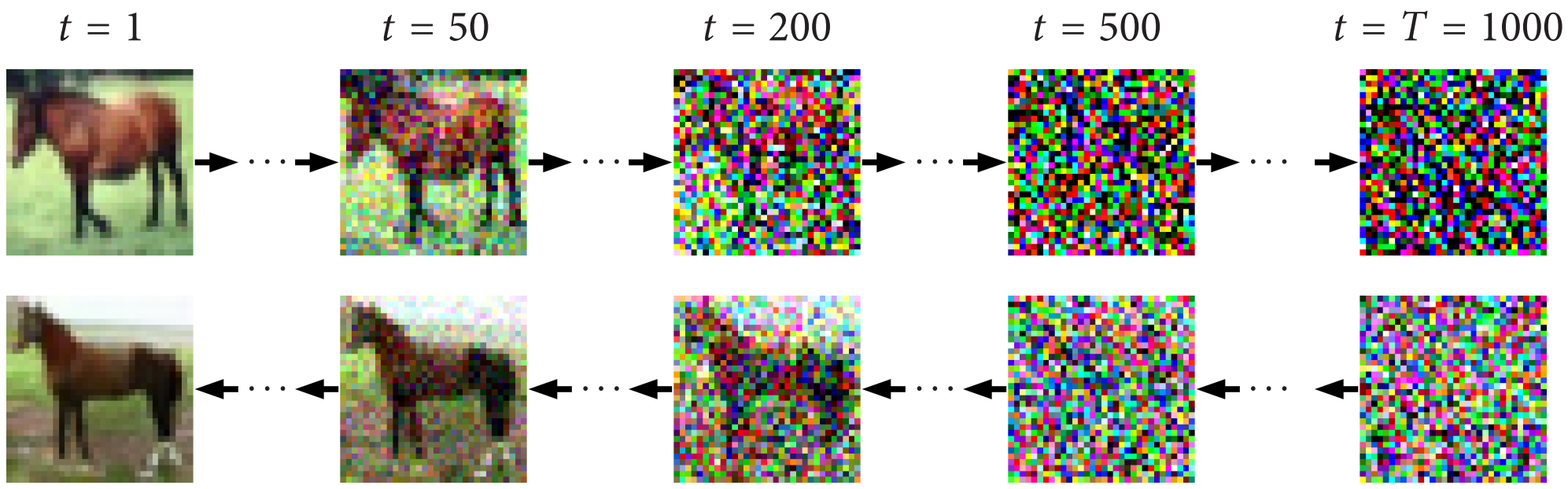

**Figure 1.7.** Samples from the forward (top) and reverse (bottom) processes (using a pretrained diffusion model for illustration; see (Ho, Jain, and Abbeel 2020)).

We introduce the forward process in the context of deriving the ELBO. Its role in denoising will become clear in Section 1.7.

**Constructing the forward diffusion process.** Since diffusion models can be viewed as latent-variable models with a hierarchical Markov structure, direct maximization of the log-likelihood is generally intractable. We instead maximize an evidence lower bound (ELBO). For a single sample $\mathbf{x}$, the ELBO is

$$\mathcal{L}(\theta, g) = \mathsf{E}_{g(\mathbf{z}^T|\mathbf{x})}\left(\ln \frac{f_\theta(\mathbf{Z}^T, \mathbf{x})}{g(\mathbf{Z}^T|\mathbf{x})}\right), \tag{1.62}$$

where $\mathbf{z}^T = (\mathbf{z}_1, \dots, \mathbf{z}_T)$ (not to be confused with the transpose of the vector $\mathbf{z}$). We wish to choose $g(\mathbf{z}^T|\mathbf{x})$ that closely approximates $f_\theta(\mathbf{z}^T|\mathbf{x})$ while making the ELBO maximization tractable. Since the conditional pdfs of the backward diffusion process in (1.60) are Gaussian, it is convenient to express the ELBO as

$$\mathcal{L}(\theta, g) = \mathsf{E}_{g(\mathbf{z}^T|\mathbf{x})}\left[\ln \frac{f(\mathbf{Z}_T)\prod_{t=2}^T f_\theta(\mathbf{Z}_{t-1}|\mathbf{Z}_t) f_\theta(\mathbf{x}|\mathbf{Z}_1)}{g(\mathbf{Z}^T|\mathbf{x})}\right]. \tag{1.63}$$

We then choose $g(\mathbf{z}^T|\mathbf{x}) = g(\mathbf{z}_T|\mathbf{x}) \prod_{t=2}^T g(\mathbf{z}_{t-1}|\mathbf{z}_t, \mathbf{x})$ such that each conditional $g(\mathbf{z}_{t-1}|\mathbf{z}_t, \mathbf{x})$

is Gaussian. To do so, we introduce the *forward diffusion process*:

$$g(\mathbf{z}_1|\mathbf{x}) = \mathrm{N}\big(\mathbf{z}_1; \sqrt{1-\beta_1}\,\mathbf{x}, \beta_1 I\big), \tag{1.64}$$

$$g(\mathbf{z}_t|\mathbf{z}_{t-1}) = \mathrm{N}\big(\mathbf{z}_t; \sqrt{1-\beta_t}\,\mathbf{z}_{t-1}, \beta_t I\big), \tag{1.65}$$

where the parameters $\beta_1, \beta_2, \ldots, \beta_T \in (0,1)$ are chosen so that the mean of $\mathbf{Z}_t$ approaches zero and its covariance matrix approaches the identity matrix as $t$ increases. We can express this model as a cascade of additive white Gaussian noise channels

$$\mathbf{Z}_1 = \sqrt{1-\beta_1}\,\mathbf{X} + \sqrt{\beta_1}\,\mathbf{V}_1, \tag{1.66}$$

$$\mathbf{Z}_t = \sqrt{1-\beta_t}\,\mathbf{Z}_{t-1} + \sqrt{\beta_t}\,\mathbf{V}_t, \ \ t = 2, \ldots, T, \tag{1.67}$$

where $\mathbf{V}_1, \ldots, \mathbf{V}_T$ are i.i.d. $\mathrm{N}(\mathbf{0}, I)$ independent of $\mathbf{X}$. If we let $\alpha_t = \prod_{\tau=1}^{t}(1-\beta_\tau)$, a straightforward calculation shows that $\mathbf{Z}_t$ can be expressed in terms of $\mathbf{X}$ as

$$\mathbf{Z}_t = \sqrt{\alpha_t}\,\mathbf{X} + \sqrt{1-\alpha_t}\,\mathbf{W}_t, \tag{1.68}$$

where $\mathbf{W}_t \sim \mathrm{N}(\mathbf{0}, I)$ (marginally, with dependence across $t$), or equivalently as

$$g(\mathbf{z}_t|\mathbf{x}) = \mathrm{N}\big(\mathbf{z}_t; \sqrt{\alpha_t}\,\mathbf{x}, (1-\alpha_t)I\big). \tag{1.69}$$

While the conditional pdfs $g(\mathbf{z}_{t-1}|\mathbf{z}_t)$, $t = 2, \ldots, T$, of the backward version of the forward process (1.65) are not Gaussian, if we condition the entire process on $\mathbf{X} = \mathbf{x}$, the resulting *conditional backward process* is of the form

$$g(\mathbf{z}^T|\mathbf{x}) = \Big(\prod_{t=2}^{T} g(\mathbf{z}_{t-1}|\mathbf{z}_t, \mathbf{x})\Big) \cdot g(\mathbf{z}_T|\mathbf{x}), \tag{1.70}$$

and the conditional pdfs $g(\mathbf{z}_{t-1}|\mathbf{z}_t, \mathbf{x})$, $t = 2, \ldots, T$, are Gaussian. In particular,

$$g(\mathbf{z}_{t-1}|\mathbf{z}_t, \mathbf{x}) = \mathrm{N}\big(\mathbf{z}_{t-1}; \mathbf{m}_t(\mathbf{x}, \mathbf{z}_t), \sigma_t^2 I\big), \tag{1.71}$$

with

$$\mathbf{m}_t(\mathbf{x}, \mathbf{z}_t) = \frac{(1-\alpha_{t-1})\sqrt{1-\beta_t}}{1-\alpha_t}\,\mathbf{z}_t + \frac{\sqrt{\alpha_{t-1}}\,\beta_t}{1-\alpha_t}\mathbf{x}, \tag{1.72}$$

$$\sigma_t^2 = \frac{\beta_t(1-\alpha_{t-1})}{1-\alpha_t}. \tag{1.73}$$

The derivation of these formulas is left as an exercise (Problem 1.4). We match the covariances of this conditional backward process to the diffusion model process in (1.60) by setting

$$\beta'_1 = \beta_1, \qquad \beta'_t = \frac{\beta_t(1-\alpha_{t-1})}{1-\alpha_t}, \ \ t = 2, \ldots, T. \tag{1.74}$$

**Optimizing the ELBO.** With the above choice of $g$, we proceed to optimizing the ELBO.

**Theorem 1.6.** The set of parameters $\hat{\theta}$ that maximizes the ELBO in (1.62) with $g(\mathbf{z}^T|\mathbf{x})$ as defined in (1.70) is given by

$$\hat{\theta} = \arg\max_{\theta} \Big[ \mathsf{E}_{g(\mathbf{z}_1|\mathbf{x})}\big(\ln f_\theta(\mathbf{x}|\mathbf{Z}_1)\big) - \sum_{t=2}^{T} \mathsf{E}_{g(\mathbf{z}_t|\mathbf{x})}\Big(\frac{1}{2\sigma_t^2}\Big\|\mathbf{m}_t(\mathbf{x},\mathbf{Z}_t) - \boldsymbol{\mu}_t(\mathbf{Z}_t)\Big\|_2^2\Big)\Big]. \tag{1.75}$$

**Proof.** Consider the ELBO in (1.62)

$$\mathcal{L}(\theta, g) = \mathsf{E}_{g(\mathbf{z}^T|\mathbf{x})}\Big(\ln \frac{f_\theta(\mathbf{Z}^T,\mathbf{x})}{g(\mathbf{Z}^T|\mathbf{x})}\Big) \tag{1.76}$$

$$= \mathsf{E}_{g(\mathbf{z}^T|\mathbf{x})}\Big(\ln \frac{f(\mathbf{Z}_T)\prod_{t=2}^{T} f_\theta(\mathbf{Z}_{t-1}|\mathbf{Z}_t) f_\theta(\mathbf{x}|\mathbf{Z}_1)}{g(\mathbf{Z}_T|\mathbf{x})\prod_{t=2}^{T} g(\mathbf{Z}_{t-1}|\mathbf{Z}_t,\mathbf{x})}\Big) \tag{1.77}$$

$$= \mathsf{E}_{g(\mathbf{z}^T|\mathbf{x})}\Big(\ln f(\mathbf{Z}_T) - \ln g(\mathbf{Z}_T|\mathbf{x}) + \ln f_\theta(\mathbf{x}|\mathbf{Z}_1) + \sum_{t=2}^{T} \ln \frac{f_\theta(\mathbf{Z}_{t-1}|\mathbf{Z}_t)}{g(\mathbf{Z}_{t-1}|\mathbf{Z}_t,\mathbf{x})}\Big) \tag{1.78}$$

$$= \mathsf{E}_{g(\mathbf{z}_T|\mathbf{x})}\big(\ln f(\mathbf{Z}_T) - \ln g(\mathbf{Z}_T|\mathbf{x})\big) + \mathsf{E}_{g(\mathbf{z}_1|\mathbf{x})}\big(\ln f_\theta(\mathbf{x}|\mathbf{Z}_1)\big) - \sum_{t=2}^{T} \mathsf{E}_{g(\mathbf{z}_t|\mathbf{x})}\Big(D\big(g(\mathbf{z}_{t-1}|\mathbf{Z}_t,\mathbf{x})\|f_\theta(\mathbf{z}_{t-1}|\mathbf{Z}_t)\big)\Big) \tag{1.79}$$

$$= \mathsf{E}_{g(\mathbf{z}_T|\mathbf{x})}\big(\ln f(\mathbf{Z}_T) - \ln g(\mathbf{Z}_T|\mathbf{x})\big) + \mathsf{E}_{g(\mathbf{z}_1|\mathbf{x})}\big(\ln f_\theta(\mathbf{x}|\mathbf{Z}_1)\big) - \sum_{t=2}^{T} \mathsf{E}_{g(\mathbf{z}_t|\mathbf{x})}\Big(\frac{1}{2\sigma_t^2}\|\mathbf{m}_t(\mathbf{x},\mathbf{Z}_t) - \boldsymbol{\mu}_t(\mathbf{Z}_t)\|_2^2\Big), \tag{1.80}$$

where the last step follows since for $t = 2, \ldots, T$, both $g(\mathbf{z}_{t-1}|\mathbf{z}_t, \mathbf{x})$ and $f_\theta(\mathbf{z}_{t-1}|\mathbf{z}_t)$ are multivariate Gaussian with the same covariance $\sigma_t^2 I$, hence their relative entropy is given by

$$D\big(g(\mathbf{z}_{t-1}|\mathbf{z}_t,\mathbf{x})\big\|f_\theta(\mathbf{z}_{t-1}|\mathbf{z}_t)\big) = \frac{1}{2\sigma_t^2}\Big\|\mathbf{m}_t(\mathbf{x},\mathbf{z}_t) - \boldsymbol{\mu}_t(\mathbf{z}_t)\Big\|_2^2. \tag{1.81}$$

The first term in (1.80) is not a function of the parameters, so we drop it from the maximization, which completes the proof of the theorem.

Note that the first term in (1.75) corresponds to the reconstruction term of the ELBO for the VAE objective in (1.58) and can be similarly evaluated by replacing the expectation with a sample average. The second term learns the reverse diffusion process by encouraging each reverse transition $f_\theta(\mathbf{z}_{t-1}|\mathbf{z}_t)$ to approximate the corresponding posterior transition $g(\mathbf{z}_{t-1}|\mathbf{z}_t, \mathbf{x})$. Since these conditional distributions are Gaussian, the resulting objective reduces to a tractable quadratic form and can also be evaluated by first replacing the expectations with sample averages.

We have defined the diffusion model and discussed the construction of the distribution used in the ELBO. We argued that maximizing the ELBO with this choice is computationally tractable. We now explain how training and sampling are performed.

**Training procedure.** The following procedure optimizes the quadratic reverse-transition matching terms in the ELBO, corresponding to $t = 2, \ldots, T$. The reconstruction term corresponding to $t = 1$ is not included in this procedure. A neural network with input $(t, \mathbf{z}_t)$, $t \in \{2, 3, \ldots, T\}$, is used to model the reverse means $\theta = \{\boldsymbol{\mu}_t(\mathbf{z}_t)\}_{t=2}^{T}$ in (1.75). Given a dataset $\{\mathbf{x}_i\}_{i=1}^{n}$ and a selection of the $\beta_t$ values:

1. Repeat until convergence:
   (a) Select a data point $\mathbf{x}$ (or a batch of data points).
   (b) Randomly select an index $t$ uniformly from $\{2, 3, \ldots, T\}$.
   (c) Randomly sample $\mathbf{w}$ from $\mathrm{N}(\mathbf{0}, I)$.
   (d) Compute $\mathbf{z}_t = \sqrt{\alpha_t}\,\mathbf{x} + \sqrt{1 - \alpha_t}\,\mathbf{w}$ (see (1.68)).
   (e) Take a gradient descent step on $\frac{1}{2\sigma_t^2}\|\mathbf{m}_t(\mathbf{x}, \mathbf{z}_t) - \boldsymbol{\mu}_t(\mathbf{z}_t)\|_2^2$ with respect to the neural network parameters.
2. Output $\hat{\theta}$.

**Sampling procedure.** Sampling from a trained diffusion model is performed by passing a noise sample through the trained backward process. First note that we can express the model (1.60) as:

$$\mathbf{Z}_{t-1} = \boldsymbol{\mu}_t(\mathbf{Z}_t) + \sqrt{\beta_t'}\,\mathbf{U}_t, \;\; \mathbf{U}_t \sim \mathrm{N}(\mathbf{0}, I).$$

1. Independently generate samples $\mathbf{z}_T$ and $\mathbf{u}_1$ from $\mathrm{N}(\mathbf{0}, I)$.
2. For $t = T, \ldots, 2$:
   (a) Compute $\boldsymbol{\mu}_t(\mathbf{z}_t)$.
   (b) Sample $\mathbf{u}_t$ from $\mathrm{N}(\mathbf{0}, I)$.
   (c) Compute $\mathbf{z}_{t-1} = \boldsymbol{\mu}_t(\mathbf{z}_t) + \sqrt{\beta_t'}\,\mathbf{u}_t$.
3. Finally compute $\mathbf{x} = \boldsymbol{\mu}_1(\mathbf{z}_1) + \sqrt{\beta_1'}\,\mathbf{u}_1$.

**Remark.** Diffusion models can be extended to discrete distributions (see Problem 1.7).

### 1.6.2 How close is $\mathcal{L}$ to $\ell$?

Recall that, for each $\mathbf{x}$, the difference between the log-likelihood and the ELBO is

$$\ell(\theta) - \mathcal{L}(\theta, g) = D\big(g(\mathbf{z}^T|\mathbf{x})\|f_\theta(\mathbf{z}^T|\mathbf{x})\big) \geq 0. \tag{1.82}$$

Thus, the ELBO is tight when the forward conditional $g(\mathbf{z}^T|\mathbf{x})$ agrees with the posterior $f_\theta(\mathbf{z}^T|\mathbf{x})$. To understand how the choice of the forward and reverse processes affects the ELBO, it is useful to average over the data-generating distribution $f(\mathbf{x})$. Define $g(\mathbf{x}, \mathbf{z}^T) = f(\mathbf{x})g(\mathbf{z}^T|\mathbf{x})$. Then

$$\mathsf{E}_f\left(\mathcal{L}(\theta, g)\right) = -h(X) - D\big(g(\mathbf{x}, \mathbf{z}^T)\|f_\theta(\mathbf{x}, \mathbf{z}^T)\big).$$

Moreover, by the chain rule for relative entropy, applied in reverse time,

$$D(g\|f_\theta) = D\big(g(\mathbf{z}_T)\|f(\mathbf{z}_T)\big) + \sum_{t=2}^{T} \mathsf{E}_{g(\mathbf{z}_t)}\left(D\big(g(\mathbf{z}_{t-1}|\mathbf{z}_t)\|f_\theta(\mathbf{z}_{t-1}|\mathbf{z}_t)\big)\right) + \mathsf{E}_{g(\mathbf{z}_1)} D\big(g(\mathbf{x}|\mathbf{z}_1)\|f_\theta(\mathbf{x}|\mathbf{z}_1)\big). \tag{1.83}$$

Here all the distributions denoted by $g$ are induced by the joint distribution $r(\mathbf{x}, \mathbf{z}^T)$. Thus, maximizing the average ELBO amounts to making the terminal distribution $g(\mathbf{z}_T)$ close to $f(\mathbf{z}_T) = \mathrm{N}(\mathbf{0}, I)$, each reverse conditional $f_\theta(\mathbf{z}_{t-1}|\mathbf{z}_t)$ close to the corresponding reverse conditional $g(\mathbf{z}_{t-1}|\mathbf{z}_t)$, and $f_\theta(\mathbf{x}|\mathbf{z}_1)$ close to $g(\mathbf{x}|\mathbf{z}_1)$. This motivates using a large number of diffusion steps $T$, each injecting only a small amount of noise $\beta_t$. With sufficient total noise, $g(\mathbf{z}_T)$ approaches $\mathrm{N}(\mathbf{0}, I)$, while for small $\beta_t$ the reverse conditionals $g(\mathbf{z}_{t-1}|\mathbf{z}_t)$ are approximately Gaussian under suitable regularity conditions. In the limit as the number of steps tends to infinity and the noise at each step becomes infinitesimal, the forward process converges to a continuous-time diffusion process (Song et al. 2021). The corresponding reverse process is also a diffusion process whose infinitesimal transitions are Gaussian (Anderson 1982).

## 1.7 SCORE-BASED MODELS AND FISHER DIVERGENCE

Score-based models take another approach to training. Rather than minimizing relative entropy—or maximizing a surrogate lower bound (ELBO)—they minimize the Fisher divergence, which measures the discrepancy between the score functions of the data-generating distribution and the model distribution.

### 1.7.1 Fisher divergence

Relative entropy, or more generally an $f$-divergence, are used to measure the discrepancy between two probability distributions. In some cases, computing this measure is intractable.

**Definition 1.10.** Given a function $E : \mathbb{R}^K \to \mathbb{R}$ such that $e^{-E(\mathbf{x})}$ is integrable, an *energy-based* (or Gibbs) pdf is defined by

$$f(\mathbf{x}) = \frac{1}{\zeta} e^{-E(\mathbf{x})}, \ \mathbf{x} \in \mathbb{R}^K, \tag{1.84}$$

where $\zeta = \int_{\mathbb{R}^K} e^{-E(\mathbf{x})}\, d\mathbf{x}$ is referred to as the *partition function* and depends only on the choice of $E$.

Energy-based distributions constitute a broad class of probability distributions that includes the exponential family. When relative entropy, or more generally $f$-divergence, is used to compare two pdfs (or pmfs), one of which is energy-based, evaluating the corresponding partition function is required. Except in simple cases, however, such as Gaussian or exponential distributions, the partition function is computationally intractable. This motivates the search for an alternative distance measure between two such distributions that avoids the need to evaluate the partition function.

Recall that the Fisher score function $s_X(x)$ is the derivative of the log-pdf with respect to $x$. We extend this definition to vectors, assuming throughout this section that all pdfs have connected support and are continuously differentiable and positive on the interior of their support.

**Definition 1.11.** Let $f(\mathbf{x})$ be a pdf satisfying the above regularity condition. The score function of $f(\mathbf{x})$ is defined by

$$s_f(\mathbf{x}) = \begin{cases} \nabla_{\mathbf{x}} \ln f(\mathbf{x}) & \text{if } f(\mathbf{x}) > 0, \\ \mathbf{0} & \text{otherwise.} \end{cases} \tag{1.85}$$

**Example 1.1** (Score function for Gaussian)**.** For $f(\mathbf{x}) = \mathrm{N}(\mathbf{x}; \boldsymbol{\mu}, \sigma^2 I)$, the score function is

$$s_f(\mathbf{x}) = -\frac{\mathbf{x} - \boldsymbol{\mu}}{\sigma^2}. \tag{1.86}$$

**Example 1.2** (Score function for energy-based pdfs)**.** Let $f(\mathbf{x})$ be a differentiable energy-based pdf as defined in (1.84). Its score function is

$$\begin{aligned} s_f(\mathbf{x}) &= \nabla_{\mathbf{x}} \ln f(\mathbf{x}) && (1.87) \\ &= \nabla_{\mathbf{x}} \big( \ln e^{-E(\mathbf{x})} - \ln \zeta \big) && (1.88) \\ &= -\nabla_{\mathbf{x}} E(\mathbf{x}) - \nabla_{\mathbf{x}} \ln \zeta && (1.89) \\ &= -\nabla_{\mathbf{x}} E(\mathbf{x}), && (1.90) \end{aligned}$$

since the partition function $\zeta$ is independent of $\mathbf{x}$. Thus, the score function $s_f(\mathbf{x})$ can be computed without evaluating the partition function.

**Lemma 1.2.** Let $f$ and $g$ be two pdfs with a common connected support $\mathscr{S}$ satisfying the regularity conditions above. If $s_f(\mathbf{x}) = s_g(\mathbf{x})$ for all $\mathbf{x} \in \mathscr{S}$, then $f(\mathbf{x}) = g(\mathbf{x})$ for all $\mathbf{x} \in \mathscr{S}$.

The proof is left as an exercise. This lemma suggests that the distance between the score functions of two pdfs would be a good measure of the distance between the pdfs themselves.

**Definition 1.12** (Fisher divergence)**.** Let $f(\mathbf{x})$ and $g(\mathbf{x})$ be two pdfs satisfying the above regularity conditions. We define the *Fisher divergence* between $f$ and $g$ as

$$D_{\mathrm{F}}(f \| g) = \frac{1}{2} \mathsf{E}_{f(\mathbf{x})} \left( \| s_f(\mathbf{X}) - s_g(\mathbf{X}) \|_2^2 \right). \tag{1.91}$$

**Theorem 1.7** (Properties of Fisher divergence)**.** The Fisher divergence between two pdfs $f$ and $g$ with common connected support $\mathscr{S} \subset \mathbb{R}^K$ satisfying the above regularity conditions has the following properties:

1. $D_F(f\|g) \geq 0$ with equality if and only if $f = g$.
2. $D_F(f\|g)$ is not symmetric in general.
3. *Scaling.* For $K = 1$, let $y = ax$, $a \neq 0$, and let $f(y)$ and $g(y)$ be the derived pdfs from $f(x)$ and $g(x)$, respectively. Then $D_F(f(y)\|g(y)) = (1/a^2)D_F(f(x)\|g(x))$.
4. *Fisher divergence and Fisher information.* For $K = 1$, let $X \sim f$ have finite mean $\mu$ and variance $0 < \sigma^2 < \infty$, and finite Fisher information. Assume that $(x - \mu)f(x) \to 0$ at both endpoints of the support of $f$. Let $g(x) = \mathrm{N}(x; \mu, \sigma^2)$. Then

$$D_F(f\|g) = \frac{1}{2}\Big(\mathcal{I}(X) - \frac{1}{\sigma^2}\Big). \tag{1.92}$$

Parts 1 and 2 follow from the definition of the Fisher divergence. The proofs of parts 3 and 4 are left as exercises.

### 1.7.2 Denoising autoencoder

Suppose $\mathbf{X}$ is drawn from an unknown and arbitrary distribution $f_{\text{data}}(\mathbf{x})$, $\mathbf{x} \in \mathbb{R}^K$, and we are given a dataset $\{\mathbf{x}_i\}_{i=1}^n$ from which we wish to learn $f_{\text{data}}(\mathbf{x})$. The *denoising autoencoder* adopts a fundamentally different learning approach from the generative models we have discussed so far. However, as we will see, diffusion model training derived via the ELBO is essentially equivalent to denoising autoencoder training.

The idea in denoising autoencoders is to corrupt $\mathbf{X}$ with independent Gaussian noise $\mathbf{Z} \sim \mathrm{N}(\mathbf{0}, \sigma^2 I)$ to obtain the noisy data

$$\mathbf{Y} = \mathbf{X} + \mathbf{Z}. \tag{1.93}$$

This induces a smoothed data distribution $f_{\text{data}}(\mathbf{y})$, whose score function is well-defined and differentiable as a consequence of Gaussian smoothing.

Instead of learning the distribution of $\mathbf{X}$ directly, we learn the distribution of the corrupted variable $\mathbf{Y}$. The motivation is that the distribution of $\mathbf{X}$ may be arbitrary and difficult to model directly, while the addition of Gaussian noise ensures that $f_{\text{data}}(\mathbf{y})$ is smooth and differentiable. This smoothed distribution can then be learned by minimizing the Fisher divergence between the noisy data distribution and a model distribution.

We model the score function of $f_{\text{data}}(\mathbf{y})$ by a neural network $s_\theta(\mathbf{y})$. When $s_\theta$ is the score function of a model pdf $f_\theta$, the corresponding score-matching objective is the Fisher divergence between $f_{\text{data}}$ and $f_\theta$. More generally, we can minimize this objective directly without requiring $s_\theta$ to be the score function of a pdf.

**Theorem 1.8.** The optimal parameters of the score function model are

$$\hat{\theta} = \arg\min_{\theta} \frac{1}{2}\,\mathsf{E}_{f_{\text{data}}(\mathbf{x},\mathbf{y})}\Big(\Big\|\frac{\mathbf{X}-\mathbf{Y}}{\sigma^2} - s_\theta(\mathbf{Y})\Big\|_2^2\Big), \tag{1.94}$$

where $s_\theta(\mathbf{y})$ is a neural network model of $\nabla_{\mathbf{y}} \ln f_{\text{data}}(\mathbf{y})$.

**Proof.** Consider the score-matching objective

$$\frac{1}{2}\,\mathsf{E}_{f_{\text{data}}(\mathbf{y})}\left(\|s_{f_{\text{data}}}(\mathbf{Y}) - s_\theta(\mathbf{Y})\|_2^2\right) \tag{1.95}$$

$$= \frac{1}{2}\,\mathsf{E}_{f_{\text{data}}(\mathbf{y})}\left(\|s_{f_{\text{data}}}(\mathbf{Y})\|_2^2\right) + \frac{1}{2}\,\mathsf{E}_{f_{\text{data}}(\mathbf{y})}\left(\|s_\theta(\mathbf{Y})\|_2^2\right) - \mathsf{E}_{f_{\text{data}}(\mathbf{y})}\left((s_{f_{\text{data}}}(\mathbf{Y}))^T s_\theta(\mathbf{Y})\right) \tag{1.96}$$

$$= c + \frac{1}{2}\,\mathsf{E}_{f_{\text{data}}(\mathbf{y})}\left(\|s_\theta(\mathbf{Y})\|_2^2\right) - \mathsf{E}_{f_{\text{data}}(\mathbf{y})}\left((s_{f_{\text{data}}}(\mathbf{Y}))^T s_\theta(\mathbf{Y})\right), \tag{1.97}$$

for some constant $c$ that does not depend on $\theta$. Consider the third term in (1.97) (assuming for simplicity that $\mathbf{X} \sim f_{\text{data}}(\mathbf{x})$),

$$\mathsf{E}_{f_{\text{data}}(\mathbf{y})}\left((s_{f_{\text{data}}}(\mathbf{Y}))^T s_\theta(\mathbf{Y})\right) = \int f_{\text{data}}(\mathbf{y}) \cdot \frac{(\nabla_{\mathbf{y}} f_{\text{data}}(\mathbf{y}))^T}{f_{\text{data}}(\mathbf{y})} s_\theta(\mathbf{y})\, d\mathbf{y} \tag{1.98}$$

$$= \int (\nabla_{\mathbf{y}} f_{\text{data}}(\mathbf{y}))^T s_\theta(\mathbf{y})\, d\mathbf{y} \tag{1.99}$$

$$= \int \left( \int f_{\text{data}}(\mathbf{x}) f_{\mathbf{Y}|\mathbf{X}}(\mathbf{y}|\mathbf{x}) \cdot \nabla_{\mathbf{y}} \ln f_{\mathbf{Y}|\mathbf{X}}(\mathbf{y}|\mathbf{x})\, d\mathbf{x} \right)^T s_\theta(\mathbf{y})\, d\mathbf{y} \tag{1.100}$$

$$= \int\int f_{\text{data}}(\mathbf{x},\mathbf{y}) \cdot (\nabla_{\mathbf{y}} \ln \mathrm{N}(\mathbf{y}-\mathbf{x}; \mathbf{0}, \sigma^2 I))^T\, s_\theta(\mathbf{y})\, d\mathbf{x}\, d\mathbf{y} \tag{1.101}$$

$$= \mathsf{E}_{f_{\text{data}}(\mathbf{x},\mathbf{y})}\left(\frac{(\mathbf{X}-\mathbf{Y})^T}{\sigma^2}\, s_\theta(\mathbf{Y})\right). \tag{1.102}$$

Hence, minimizing the Fisher divergence between the noisy data distribution and the model reduces to a denoising score matching objective of the form

$$\frac{1}{2}\,\mathsf{E}_{f_{\text{data}}(\mathbf{y})}\left(\|s_\theta(\mathbf{Y})\|_2^2\right) - \mathsf{E}_{f_{\text{data}}(\mathbf{y})}\left((s_{f_{\text{data}}}(\mathbf{Y}))^T s_\theta(\mathbf{Y})\right) \tag{1.103}$$

$$= \frac{1}{2}\,\mathsf{E}_{f_{\text{data}}(\mathbf{x},\mathbf{y})}\left(\|s_\theta(\mathbf{Y})\|_2^2\right) - \mathsf{E}_{f_{\text{data}}(\mathbf{x},\mathbf{y})}\left(\frac{(\mathbf{X}-\mathbf{Y})^T}{\sigma^2} s_\theta(\mathbf{Y})\right) \tag{1.104}$$

$$= \frac{1}{2}\,\mathsf{E}_{f_{\text{data}}(\mathbf{x},\mathbf{y})}\left(\left\|\frac{(\mathbf{X}-\mathbf{Y})}{\sigma^2} - s_\theta(\mathbf{Y})\right\|_2^2\right) - c', \tag{1.105}$$

where $c'$ does not depend on $\theta$, which completes the proof of the theorem.

Hence, training a denoising autoencoder reduces to minimizing the mean-squared error between the score function of $f_{\text{data}}(\mathbf{y})$ (modeled by a neural network) and the score function of the Gaussian noise $-\dfrac{\mathbf{Y}-\mathbf{X}}{\sigma^2}$.

**Connection to denoising.** A classical problem in signal processing is to estimate a signal $\mathbf{X}$ from observations $\mathbf{Y}$ corrupted by additive Gaussian noise. The observation model is

$$\mathbf{Y} = \mathbf{X} + \mathbf{Z}, \tag{1.106}$$

where $\mathbf{X}$ is the signal with unknown pdf $f(\mathbf{x})$, $\mathbf{Z} \sim \mathrm{N}(\mathbf{0}, \sigma^2 I)$ is independent of $\mathbf{X}$, and all vectors have the same dimension. For example, $\mathbf{X}$ may be an image, and $\mathbf{Y}$ a noisy version of it. Given $\mathbf{Y}$, we wish to estimate $\mathbf{X}$, or equivalently to estimate the noise $\mathbf{Z}$ and subtract it from $\mathbf{Y}$. Using the mean-squared error criterion, we seek to minimize $\mathsf{E}\left(\|\mathbf{X} - \hat{x}(\mathbf{Y})\|_2^2\right)$.

It is well-known that the minimum mean-squared estimate is the conditional expectation $\mathsf{E}(\mathbf{X}|\mathbf{Y})$. The following remarkable formula relates this estimate to the score function of $f(\mathbf{y})$, which is differentiable due to Gaussian convolution.

**Theorem 1.9** (Tweedie's formula)**.** Consider the model in (1.106). The minimum mean-squared error estimate of $\mathbf{X}$ given the observation $\mathbf{Y}$ is

$$\mathsf{E}(\mathbf{X}|\mathbf{Y}=\mathbf{y}) = \mathbf{y} + \sigma^2 \nabla_{\mathbf{y}} \ln f(\mathbf{y}). \tag{1.107}$$

The proof of this theorem is left as an exercise (Problem 1.12). Since $\mathbf{Z} = \mathbf{Y} - \mathbf{X}$, Tweedie's formula can be equivalently written as $\mathsf{E}(\mathbf{Z}|\mathbf{Y}=\mathbf{y}) = -\sigma^2 \nabla_{\mathbf{y}} \ln f(\mathbf{y})$. Thus, minimizing the mean-squared error in estimating the signal $\mathbf{X}$ is equivalent to minimizing the mean-squared error in estimating the noise $\mathbf{Z}$, and the optimal noise estimator is proportional to the score function of the noisy observation $\mathbf{Y}$.

### 1.7.3 Diffusion model training as denoising score matching

In Section 1.6, we showed that the main component of the diffusion model training objective is a sum of quadratic reverse-transition matching terms. We now show that these terms can be interpreted as a weighted sum of denoising score-matching objectives at different noise levels.

Recall the second term in the training objective for diffusion models (1.75):

$$\sum_{t=2}^{T} \mathsf{E}_{g(\mathbf{z}_t|\mathbf{x})}\Big(\frac{1}{2\sigma_t^2}\left\|\mathbf{m}_t(\mathbf{x},\mathbf{Z}_t) - \boldsymbol{\mu}_t(\mathbf{Z}_t)\right\|_2^2\Big). \tag{1.108}$$

From Tweedie's formula,

$$\mathbf{m}_t(\mathbf{x},\mathbf{z}_t) = \mathsf{E}(\mathbf{Z}_{t-1}|\mathbf{Z}_t=\mathbf{z}_t, \mathbf{X}=\mathbf{x}) = \frac{1}{\sqrt{1-\beta_t}}\Big(\mathbf{z}_t + \beta_t \nabla_{\mathbf{z}_t} \ln g(\mathbf{z}_t|\mathbf{x})\Big). \tag{1.109}$$

Motivated by this representation, we parametrize the mean of the reverse transition as

$$\boldsymbol{\mu}_t(\mathbf{z}_t) = \mathsf{E}_{f_\theta(\mathbf{z}_{t-1}|\mathbf{z}_t)}(\mathbf{Z}_{t-1}) = \frac{1}{\sqrt{1-\beta_t}}\Big(\mathbf{z}_t + \beta_t s_\theta(\mathbf{z}_t, t)\Big). \tag{1.110}$$

Hence, the $t$th quadratic term in (1.108) becomes

$$\frac{\beta_t^2}{2\sigma_t^2(1-\beta_t)}\,\mathsf{E}_{g(\mathbf{z}_t|\mathbf{x})}\Big(\left\|\nabla_{\mathbf{z}_t} \ln g(\mathbf{Z}_t|\mathbf{x}) - s_\theta(\mathbf{Z}_t,t)\right\|_2^2\Big). \tag{1.111}$$

Since $g(\mathbf{z}_t|\mathbf{x}) = \mathrm{N}\big(\mathbf{z}_t;\, \sqrt{\alpha_t}\mathbf{x},\, (1-\alpha_t)I\big)$, we have

$$\nabla_{\mathbf{z}_t} \ln g(\mathbf{z}_t|\mathbf{x}) = \frac{\sqrt{\alpha_t}\mathbf{x} - \mathbf{z}_t}{1-\alpha_t}.$$

Substituting in (1.108), we obtain

$$\sum_{t=2}^{T} \frac{\beta_t^2}{2\sigma_t^2(1-\beta_t)} \mathsf{E}_{g(\mathbf{z}_t|\mathbf{x})}\left(\left\|\frac{\sqrt{\alpha_t}\mathbf{x}-\mathbf{Z}_t}{1-\alpha_t} - s_\theta(\mathbf{Z}_t, t)\right\|_2^2\right). \tag{1.112}$$

Thus, minimizing (1.108) amounts to minimizing a weighted sum of denoising score-matching objectives, one for each noise level.

To relate this objective to Fisher divergence, let $g_t(\mathbf{z})$ denote the marginal pdf of $\mathbf{Z}_t$ in the forward diffusion process. Applying Theorem 1.8, minimizing

$$\mathsf{E}\left(\left\|\frac{\sqrt{\alpha_t}\mathbf{X}-\mathbf{Z}_t}{1-\alpha_t} - s_\theta(\mathbf{Z}_t, t)\right\|_2^2\right)$$

over $\theta$ is equivalent, up to a term independent of $\theta$, to minimizing

$$\mathsf{E}_{g_t(\mathbf{z})}\left(\|\nabla_{\mathbf{z}} \ln g_t(\mathbf{Z}) - s_\theta(\mathbf{Z}, t)\|_2^2\right).$$

Thus, each denoising score-matching objective in (1.112) is equivalent to matching $s_\theta(\mathbf{z}, t)$ to the score of $g_t$ in squared $L_2$ distance. Hence, the main component of the diffusion model training objective can be viewed as minimizing a weighted sum of score-matching objectives at different noise levels.

## 1.8 REGULARIZATION AND GENERALIZATION

Complex statistical learning models can often fit the training data extremely well, even in the presence of noise or when the true input–output relationship is substantially simpler than the model. In such cases, the model may achieve very low training error while performing poorly on unseen data, a phenomenon known as *overfitting*.

This tension between fitting the training data and generalizing to unseen data arises throughout statistical learning, from classical regression and classification models to deep generative models.

To improve generalization performance, the training procedure is often modified to discourage overly complex solutions. Techniques designed to achieve this are collectively referred to as *regularization*.

One of the simplest and most widely used forms of regularization penalizes large parameter values during training. Instead of maximizing, for example, the log-likelihood $\ell(\theta)$ alone, one optimizes a regularized objective of the form $\ell(\theta) - \lambda R(\theta)$, where $R(\theta)$ measures the complexity of the parameter vector and $\lambda > 0$ controls the strength of the regularization. Common choices include the squared Euclidean norm $R(\theta) = \|\theta\|_2^2$, known as $L_2$ *regularization*, and the $L_1$ norm $R(\theta) = \|\theta\|_1$, which encourages sparse parameter vectors.

This form of regularization can be interpreted through the lens of the minimum description length (MDL) principle. In MDL, one seeks a model that achieves a balance between goodness of fit and model complexity. Regularized learning objectives capture the same tradeoff: the log-likelihood term rewards accurate fitting of the training data, while the regularization term penalizes overly complex parameter choices that may lead to overfitting. From this perspective, the negative log-likelihood corresponds to the description length of the data given the model, whereas the regularization term can be viewed as a complexity penalty representing the description length of the model itself.

In deep neural networks, it has been observed that learned models can perfectly fit the training data, while still generalizing well to unseen inputs (Neyshabur, Tomioka, and Srebro 2015). Remarkably, this behavior can occur even in the absence of explicit regularization, suggesting that regularization may arise implicitly from algorithmic choices such as stochastic gradient descent and early stopping, in which training is terminated before convergence to a local minimum.

## SUMMARY

**Supervised learning.** Given a dataset $\{(\mathbf{x}_i, \mathbf{y}_i)\}_{i=1}^n$, supervised learning aims to find a conditional model $p_\theta(\mathbf{y}|\mathbf{x})$, $\theta \in \boldsymbol{\Theta}$, that minimizes the conditional cross entropy $H\big(p_{\text{data}}(\mathbf{y}|\mathbf{x}), p_\theta(\mathbf{y}|\mathbf{x})|p_{\text{data}}(\mathbf{x})\big)$. This is asymptotically achieved by maximizing the conditional log-likelihood $\ell_n(\theta) = \sum_{i=1}^n \ln p_\theta(\mathbf{y}_i|\mathbf{x}_i)$.

**Autoregressive models.** $p(\mathbf{x}) = p(x_1)p(x_2|x_1)\ldots p(x_K|x^{K-1})$. Training is performed using a single neural network that models all conditional distributions.

**Latent variable models.** $p(\mathbf{z}, \mathbf{x})$ with only $\mathbf{x}$ observed.

**ELBO.** For any pdf $g(\mathbf{z})$, the log-likelihood is lower bounded as

$$\ln f_\theta(\mathbf{x}) \ge \mathsf{E}_{g(\mathbf{z})}\big(\ln f_\theta(\mathbf{x}|\mathbf{Z})\big) - D\big(g(\mathbf{z})\| f_\theta(\mathbf{z})\big).$$

**Variational autoencoder.** $f(\mathbf{z}) = \mathrm{N}(\mathbf{z}; \mathbf{0}, I)$, $f_\theta(\mathbf{x}|\mathbf{z}) = \mathrm{N}(\mathbf{x}; \boldsymbol{\mu}_\theta(\mathbf{z}), \Sigma_\theta(\mathbf{z}))$. Trained using ELBO by simultaneously optimizing an encoder $g_\phi(\mathbf{z}|\mathbf{x})$ and a decoder $f_\theta(\mathbf{x}|\mathbf{z})$.

**Diffusion models.** Defined by backward diffusion process

$$\begin{aligned} f(\mathbf{z}_T) &= \mathrm{N}(\mathbf{z}_T; \mathbf{0}, I), \\ f_\theta(\mathbf{z}_{t-1}|\mathbf{z}_t) &= \mathrm{N}\big(\mathbf{z}_{t-1}; \boldsymbol{\mu}_t(\mathbf{z}_t), \beta_t' I\big),\ \ t = T, \ldots, 2, \\ f_\theta(\mathbf{x}|\mathbf{z}_1) &= \mathrm{N}(\mathbf{x}; \boldsymbol{\mu}_1(\mathbf{z}_1), \beta_1' I). \end{aligned}$$

Trained by maximizing ELBO using a fixed forward diffusion process.

**Fisher divergence.** For differentiable pdfs $f(\mathbf{x})$ and $g(\mathbf{x})$:

$$D_{\mathrm{F}}(f\|g) = \frac{1}{2}\int f(\mathbf{x})\Big(\|\nabla \ln f(\mathbf{x}) - \nabla \ln g(\mathbf{x})\|_2^2\Big)\, d\mathbf{x}.$$

**Score-based models.** Trained by minimizing the Fisher divergence.

**Denoising autoencoders.** $\mathbf{X}$ has arbitrary distribution. Denoise $\mathbf{Y} = \mathbf{X} + \mathbf{Z}$, where $\mathbf{Z} \sim \mathrm{N}(\mathbf{0}, \sigma^2 I)$ is independent of $\mathbf{X}$.

**Tweedie's formula.** Let $\mathbf{Y} = \mathbf{X} + \mathbf{Z}$, where $\mathbf{Z} \sim \mathrm{N}(\mathbf{0}, \sigma^2 I)$. Then $\mathsf{E}(\mathbf{X}|\mathbf{Y} = \mathbf{y}) = \mathbf{y} + \sigma^2 \nabla_{\mathbf{y}} \ln f(\mathbf{y})$. The main quadratic component of ELBO-based diffusion-model training is equivalent to a weighted sum of denoising score-matching objectives.

## PROBLEMS

**1.1.** *Prediction with logarithmic loss and the value of side information.* Let $(X, Y)$ be random variables taking values in a discrete set $\mathcal{X} \times \mathcal{Y}$. Suppose a predictor observes $Y = y$ and wishes to predict $X$. The Hamming distortion measure (0-1 loss function) makes a hard decision by assigning 1 if the prediction is not equal to the source symbol and 0 otherwise. The log loss function, which is widely used in machine learning, makes "soft decisions" by assigning probabilities $q_y(x)$ to the source symbols, where $q_y(x)$ is selected from the set of pmfs $\mathscr{P}(\mathcal{X})$ on $\mathcal{X}$:

$$q_y(x) \geq 0, \qquad \sum_{x \in \mathcal{X}} q_y(x) = 1.$$

The predictor's true conditional belief about $X$, given $Y = y$, is $p_{X|Y}(x|y)$, but the reported distribution $q_y$ need not equal this conditional distribution.

If the outcome is $X = x$, the predictor incurs the loss given by the surprise

$$\ell(x, q_y) = \ln\left(\frac{1}{q_y(x)}\right).$$

In the following assume that $H(X|Y) < \infty$.

(a) Prove the decomposition

$$\mathsf{E}\left(\ln \frac{1}{q_Y(X)}\right) = H(X|Y) + \mathsf{E}_Y\left(D\left(p_{X|Y}(x|Y) \,\|\, q_Y(x)\right)\right).$$

Deduce that

$$\inf_q \mathsf{E}\left(\ln \frac{1}{q_Y(X)}\right) = H(X|Y),$$

and show that the infimum is attained if and only if $q_y(x) = p_{X|Y}(x|y)$, $x \in \mathcal{X}$, for every $y$ such that $p_Y(y) > 0$. Thus, conditional entropy has a one-shot operational interpretation: it is the minimum expected logarithmic loss for predicting $X$ after observing $Y$; no block coding or asymptotic limit is required. This complements the asymptotic operational meaning of the same quantity in the rate–distortion problem with log loss.

(b) Suppose that the predictor does not observe $Y$. Show that the minimum expected logarithmic loss is $H(X)$. Conclude that the reduction in the optimal expected loss obtained by observing $Y$ is $H(X) - H(X|Y) = I(X; Y)$.

This problem is due to A. Gohari.

**1.2.** *Concavity of the log-likelihood for logistic regression.* Show that the log-likelihood $\ell(\mathbf{w})$ for logistic regression is concave in $\mathbf{w}$.

**1.3.** *Perplexity and natural language processing.* Let $p$ and $q$ be two pmfs on a finite set $\mathcal{X}$. We define the perplexity of $p$ with respect to $q$ as

$$\text{Perplexity}(p, q) = e^{H(p,q)},$$

where $H(p, q) = -\sum_x p(x) \ln q(x)$ is the cross entropy. If $p = q$, then $\text{Perplexity}(p, p) = e^{H(p)}$.

Now consider an autoregressive model (1.37) trained on a dataset $\{\mathbf{x}_i\}_{i=1}^n$, where $\mathbf{x}_i \in \mathcal{X}^K$ and $|\mathcal{X}|$ is the vocabulary size (i.e., the number of words or tokens). The performance is evaluated by calculating the empirical perplexity, defined as the exponential of the average negative log-likelihood per token:

$$\text{Perplexity}(\hat{\theta}) = \left( \prod_{i=1}^{n} \prod_{k=2}^{K} \frac{1}{p_{\hat{\theta}}(x_{ki} | x_{1i}^{k-1})} \right)^{1/(n(K-1))}.$$

(a) We know that $H(p)$ is the minimum expected description length of the random variable $X \sim p$ (within $\ln 2$ nats). How would you interpret $\text{Perplexity}(p, p)$? Provide an example.

(b) For a fixed true distribution $p$, what model distribution $q$ minimizes the perplexity $\text{Perplexity}(p, q)$?

(c) The empirical perplexity is calculated on a finite test set. Under what conditions would this empirical value be a good approximation of the true per-token perplexity

$$\exp\left( \frac{1}{K-1} \mathsf{E}\left[ -\sum_{k=2}^{K} \ln p_\theta(X_k | X^{k-1}) \right] \right)?$$

(d) Explain why low empirical perplexity indicates a high-quality language model. Specifically, how does it relate to the model's ability to predict the next token in a sequence?

The notion of perplexity first appeared in Jelinek et al. (1977) and was recently adopted as a measure of large language model performance.

**1.4.** *Conditional backward process.* Show that the conditional backward process in (1.70) is of the form

$$\frac{g(\mathbf{x}, \mathbf{z}^T)}{f(\mathbf{x})} = g(\mathbf{z}_T | \mathbf{x}) \prod_{t=2}^{T} g(\mathbf{z}_{t-1} | \mathbf{z}_t, \mathbf{x}),$$

where $g(\mathbf{x}, \mathbf{z}^T) = f(\mathbf{x}) g(\mathbf{z}^T | \mathbf{x})$, $g(\mathbf{z}_{t-1} | \mathbf{z}_t, \mathbf{x}) = \mathrm{N}(\mathbf{z}_{t-1}; \mathbf{m}_t(\mathbf{x}, \mathbf{z}_t), \sigma_t^2 I)$ with $\mathbf{m}_t(\mathbf{x}, \mathbf{z}_t)$ and $\sigma_t^2$ as given in (1.72) and (1.73), respectively.

**1.5.** *Is the forward diffusion process entropy increasing?* Consider the forward diffusion process in (1.66) and (1.67) with $\mathbf{X}$ having zero mean and covariance matrix $I$. Show that $h(\mathbf{Z}_t) \geq h(\mathbf{Z}_{t-1})$ for $t = 2, \ldots, T$. Hint: Show that $\Sigma_{Z_t} = I$ for $t = 1, 2, \ldots, T$ and use the entropy power inequality.

**1.6.** *Learning the noise in diffusion models.* Consider the diffusion model training. Parameterize the reverse mean in terms of a noise-prediction network $\boldsymbol{\nu}_t$ as

$$\boldsymbol{\mu}_t(\mathbf{z}_t) = \frac{1}{\sqrt{1-\beta_t}}\left(\mathbf{z}_t - \frac{\beta_t}{\sqrt{1-\alpha_t}}\boldsymbol{\nu}_t(\mathbf{z}_t)\right).$$

Show that

$$\begin{aligned} &D\big(g(\mathbf{z}_{t-1}|\mathbf{z}_t, \mathbf{x}) \| f_\theta(\mathbf{z}_{t-1}|\mathbf{z}_t)\big) \\ &\quad = \frac{\beta_t}{2(1-\alpha_{t-1})(1-\beta_t)}\left\|\boldsymbol{\nu}_t\big(\sqrt{\alpha_t}\,\mathbf{x} + \sqrt{1-\alpha_t}\,\mathbf{w}_t\big) - \mathbf{w}_t\right\|_2^2 + c_1, \quad t = 2, \ldots, T, \\ &\ln f_\theta(\mathbf{x}|\mathbf{z}_1) = -\frac{1}{2(1-\beta_1)}\|\boldsymbol{\nu}_1(\mathbf{z}_1) - \mathbf{w}_1\|_2^2 + c_2, \end{aligned}$$

where $c_1, c_2$ are constants that do not depend on the model parameters. Hence, training the diffusion model is equivalent to estimating the noise $\mathbf{w}_t$, $t = 1, 2, \ldots, T$.

**1.7.** *Discrete diffusion models.* In this problem we explore a discrete counterpart of the Gaussian diffusion model. Let $X$ be a discrete random variable taking values in $\{1, \ldots, m\}$. The generative discrete diffusion process is defined by the backward Markov process

$$\begin{aligned} p(z_T) &= \mathrm{Unif}(z_T; \{1, \ldots, m\}), \\ p_\theta(z_{t-1}|z_t) &= \mathrm{Cat}(z_{t-1}; \mathbf{p}_\theta(z_t)), \ t = 2, \ldots, T, \\ p_\theta(x|z_1) &= \mathrm{Cat}(x; \mathbf{p}_\theta(z_1)), \end{aligned}$$

where $\mathbf{p}_\theta(z_t)$ is an $m$-dimensional probability vector specified by a neural network with parameter $\theta$. To train this model from an observation $x$ of $X$, consider the auxiliary forward process

$$\begin{aligned} q(z_1|x) &= \beta_1 \mathbb{1}_{\{z_1 = x\}} + \frac{1}{m}(1-\beta_1), \\ q(z_t|z_{t-1}) &= \beta_t \mathbb{1}_{\{z_t = z_{t-1}\}} + \frac{1}{m}(1-\beta_t), \quad t = 2, \ldots, T, \ 0 \leq \beta_t \leq 1 - \epsilon, \ \epsilon > 0. \end{aligned}$$

Let $\alpha_t = \prod_{s=1}^t \beta_s$.

(a) Show that $\alpha_t \to 0$ as $t \to \infty$.

(b) Show that $q(z_t|x) = \alpha_t \mathbb{1}_{\{z_t = x\}} + \frac{1}{m}(1-\alpha_t)$. Conclude that $q(z_t|x) \to \frac{1}{m}$ for every $z_t$, independent of $x$.

(c) Show that

$$q(z_{t-1}|z_t, x) = \frac{q(z_t|z_{t-1})q(z_{t-1}|x)}{q(z_t|x)}$$

and express it explicitly in terms of $\beta_t$, $\alpha_{t-1}$, $x$, $z_{t-1}$, and $z_t$.

(d) Show that, up to terms that do not depend on $\theta$, maximizing the ELBO using the auxiliary forward process is equivalent to minimizing

$$\sum_{t=2}^{T} \mathsf{E}_{q(z_t|x)} \Big( D\big(q(z_{t-1}|Z_t, x) \| p_\theta(z_{t-1}|Z_t)\big)\Big) - \mathsf{E}_{q(z_1|x)} \big( \ln p_\theta(x|Z_1)\big).$$

(e) Show that the minimization reduces further to minimizing categorical cross-entropy losses, which is analogous to the quadratic losses that arise in the Gaussian case.

This problem is based on (Hoogeboom et al. 2021).

**1.8.** *Energy-based models.* Consider the scalar ($K = 1$) energy-based model

$$f_\theta(x) = \frac{1}{\zeta_\theta} e^{-E_\theta(x)}, \quad \zeta_\theta = \int e^{-E_\theta(x)}\, dx, \; \theta \in \mathbb{R},$$

where $\theta$ ranges over the values for which $0 < \zeta_\theta < \infty$.

(a) Show that at any interior value of $\theta$ for which the required derivatives exist,

$$\frac{\partial}{\partial\theta} D(f_{\text{data}} \| f_\theta) = -\mathsf{E}_{f_\theta} \Big( \frac{\partial}{\partial\theta} E_\theta(X)\Big) + \mathsf{E}_{f_{\text{data}}} \Big( \frac{\partial}{\partial\theta} E_\theta(X)\Big).$$

Assume that in all terms, integrals exist and you can interchange differentiation and integration.

(b) Let $E_\theta(x) = -\theta T(x)$. Show that any stationary point of $D(f_{\text{data}} \| f_\theta)$ satisfies $\mathsf{E}_{f_\theta}[T(X)] = \mathsf{E}_{f_{\text{data}}}[T(X)]$.

(c) Assume that $f_{\text{data}}$ and $E_\theta$ are sufficiently smooth and that

$$f_{\text{data}}(x) \frac{\partial}{\partial x} E_\theta(x) \to 0$$

at both endpoints of the support of $f_{\text{data}}$. Show that for the energy-based model, minimizing the Fisher divergence over $\theta$ is equivalent to minimizing

$$\mathsf{E}_{f_{\text{data}}} \Big[ \Big( \frac{\partial}{\partial x} E_\theta(X)\Big)^2 - 2 \frac{\partial^2}{\partial x^2} E_\theta(X)\Big].$$

**1.9.** *Fisher divergence.* Prove Lemma 1.2.

**1.10.** *Properties of Fisher divergence.* Prove part 3 of Theorem 1.7.

**1.11.** *Fisher divergence and Fisher information.* Prove part 4 of Theorem 1.7.

**1.12.** *Proof of Tweedie's formula.* Prove the scalar version of Theorem 1.9 for which $Y = X + Z$, where $Z \sim \mathrm{N}(0, \sigma^2)$ and $X$ has an unknown distribution.

**1.13.** *Generative adversarial network (GAN).* The GAN is a latent variable model $f(\mathbf{z})p_\theta(\mathbf{x}|\mathbf{z})$, where $f(\mathbf{z}) = \mathrm{N}(\mathbf{z}; \mathbf{0}, I)$ and $p_\theta(\mathbf{x}|\mathbf{z})$ is induced by a deterministic mapping $\mathbf{z} \mapsto \mathbf{x} = g_\theta(\mathbf{z})$ called the *generator*, modeled by a neural network. Training a GAN uses the two-player minimax game illustrated below. The first player trains $g_\theta(\mathbf{z})$ to generate samples that closely resemble the data, while the second player trains another neural network $d_\phi(\mathbf{x}) \in [0, 1]$, called the *discriminator*, to try to distinguish between real data and the fake data produced by $g_\theta(\mathbf{z})$.

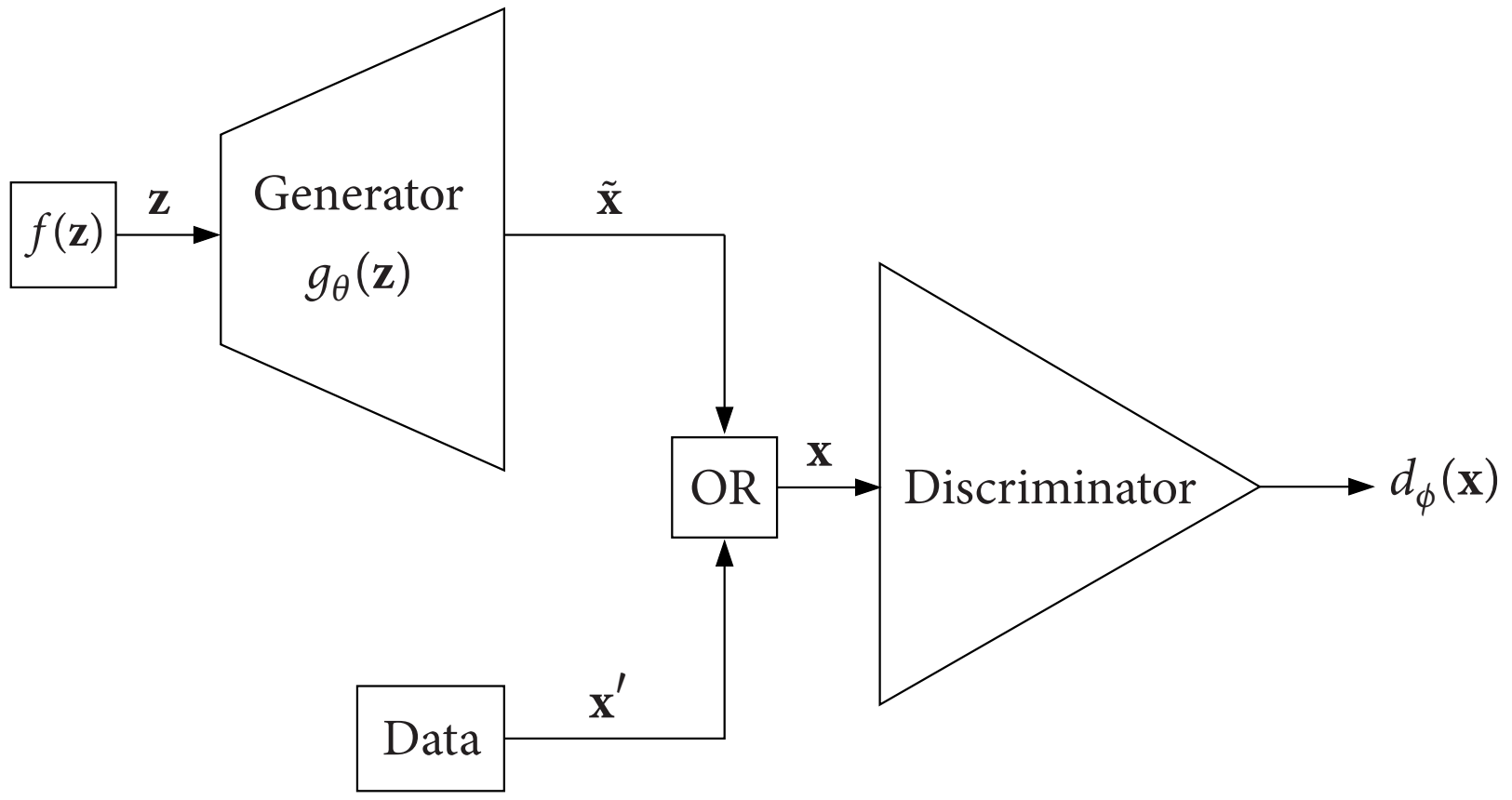

For a fixed generator $g_\theta$, which induces a distribution $p_\theta(\mathbf{x})$, the discriminator is trained to maximize the value function (or minimize the classification loss):

$$V(g_\theta, d_\phi) = \mathsf{E}_{p_{\mathrm{data}}(\mathbf{x})}\left[\ln d_\phi(\mathbf{X})\right] + \mathsf{E}_{p_\theta(\mathbf{x})}\left[\ln(1 - d_\phi(\mathbf{X}))\right],$$

where the first term corresponds to classifying the input as true when it is a data point and the second corresponds to classifying it as false when it is a fake input. Maximizing the sum improves the discriminator's ability to correctly classify fake versus true data.

(a) Show that the discriminator that maximizes $V(g_\theta, d)$ is $d^*(\mathbf{x}) = \dfrac{p_{\mathrm{data}}(\mathbf{x})}{p_{\mathrm{data}}(\mathbf{x}) + p_\theta(\mathbf{x})}$ when $p_{\mathrm{data}}(\mathbf{x}) + p_\theta(\mathbf{x}) > 0$, and may be assigned an arbitrary value in $[0, 1]$ otherwise.

(b) Substitute $d^*$ into the value function and show that

$$\max_d V(g_\theta, d) = 2(\ln 2) D_{\mathrm{JS}}(p_{\mathrm{data}} \| p_\theta) - \ln 4,$$

where $D_{\mathrm{JS}}(p\|q)$ is the Jensen–Shannon divergence.

(c) Conclude that the minimum of $\max_d V(g_\theta, d)$ is attained when $p_\theta = p_{\mathrm{data}}$, provided the generator family is rich enough to realize $p_{\mathrm{data}}$. What is the optimal discriminator on the support of $p_{\mathrm{data}}$, and what is the value of the minimax objective in this case?

Generative adversarial networks were introduced by Goodfellow et al. (2014).

**1.14.** *$f$-GAN.* Show that optimizing each of the following alternative value functions in a manner similar to the previous problem leads to a supremum that is also an $f$-divergence, up to an additive constant.

(a) $V(g_\theta, d_\phi) = \mathsf{E}_{p_{\text{data}}(\mathbf{x})}\left(d_\phi(\mathbf{X})\right) - \gamma\, \mathsf{E}_{p_\theta(\mathbf{x})}\left(d_\phi(\mathbf{X})\right)$, restricted to $|d_\phi(\mathbf{x})| \le \frac{1}{2}$, $\gamma \ge 1$.

(b) $V(g_\theta, d_\phi) = \mathsf{E}_{p_{\text{data}}(\mathbf{x})}\left(d_\phi(\mathbf{X})\right) - \mathsf{E}_{p_\theta(\mathbf{x})}\left((1/4)d_\phi^2(\mathbf{X}) + d_\phi(\mathbf{X})\right)$.

(c) $V(g_\theta, d_\phi) = \mathsf{E}_{p_{\text{data}}(\mathbf{x})}\left(d_\phi(\mathbf{X})\right) - \mathsf{E}_{p_\theta(\mathbf{x})}\left(\dfrac{d_\phi(\mathbf{X})}{1 - d_\phi(\mathbf{X})}\right)$ restricted to $d_\phi(\mathbf{x}) < 1$.

This problem is based on (Nowozin, Cseke, and Tomioka 2016).

## BIBLIOGRAPHIC NOTES

Least squares was first introduced by Legendre (1805), then by Gauss (1809) who mentioned that he had invented it in 1795. Logistic regression first appeared in (Cox 1958).

Neural networks started with the biological neuron model by McCulloch and Pitts (1943). Rosenblatt (1958) introduced the perceptron. Modern neural networks were introduced in the 1980s. The backpropagation algorithm was proposed in an early form by Werbos (1974) and subsequently developed independently by several researchers, including LeCun (1985) and Rumelhart, Hinton, and Williams (1986). The earliest universal approximation results for two-layer neural networks were established independently by Cybenko (1989), Hornik, Stinchcombe, and White (1989), and Funahashi (1989). These results were generalized by Leshno et al. (1993), who characterized precisely which activation functions yield universal approximation, and have since been extended to deep neural networks; see, for example, (Hanin and Sellke 2017). Hochreiter and Schmidhuber (1997) introduced the Long Short-Term Memory (LSTM), which enabled major developments in natural language processing. Krizhevsky, Sutskever, and Hinton (2012) marked the first breakthrough in the application of deep neural networks to computer vision. Vaswani et al. (2017) introduced the transformer model, which became the foundation for large language models.

PCA was first introduced by Pearson (1901) and later formalized by Hotelling (1929). The probabilistic version of PCA we discussed is due to Tipping and Bishop (1999). For a review of the ELBO and variational inference, see Jordan et al. (1999). The variational autoencoder was first introduced in Kingma and Welling (2013) and further developed in Rezende, Mohamed, and Wierstra (2014). Generative diffusion models were first introduced by Sohl-Dickstein et al. (2015).

Denoising autoencoders were first introduced by Vincent et al. (2008). Theorem 1.8 is due to Vincent (2011). Ho, Jain, and Abbeel (2020) introduced the denoising diffusion model and demonstrated its connection to score-based models. Hyvärinen (2005) introduced the score matching objective for training energy-based models. Tweedie's formula appears in the work of Efron (2011), who notes—following Robbins (1956)—that the result is due to Maurice Tweedie.

Our coverage of statistical learning focused primarily on some of the objective functions used in learning, leaving the details of training and implementation as well as many other types of models and techniques to specialized books on the subject, such as Hastie, Tibshirani, and Friedman (2009), Goodfellow, Bengio, and Courville (2016), Murphy (2023), and Bishop and Bishop (2024). The chapter was significantly shaped by the unpublished class notes of Ermon (2023) and the book of Bishop and Bishop (2024).

The chapter focused on models trained using supervised and unsupervised learning. Another very important, but conceptually distinct paradigm, is *reinforcement learning*. In this setting, learning takes place through sequential interaction with an environment. At each time step, an agent selects an action based on its current state, receives a reward, and transitions to a new state according to an unknown probability distribution. The objective is to learn a policy for choosing actions so as to maximize a long-term cumulative reward. Unlike supervised learning, there is no explicit target output for each input; instead, feedback is indirect, delayed, and typically noisy. Connections between reinforcement learning and information theory can be found in (Lu et al. 2024).

Statistical learning has found applications across nearly every domain, including conversational systems based on large language models such as GPT-4 (OpenAI 2023), code generation (Chen et al. 2021), computer vision (He et al. 2016), medical diagnosis (Esteva et al. 2017), scientific discovery, including the prediction of three-dimensional protein structures from amino acid sequences (Jumper et al. 2021), autonomous driving (Hwang et al. 2024), and robotics (Firoozi et al. 2024).